\documentclass[acmsmall,screen,nonacm]{acmart}

\usepackage{tabularx}
\usepackage{booktabs}
\usepackage{tikz}
\usetikzlibrary{arrows.meta}

\usepackage{hyperref}
\usepackage{hyperxmp}
\usepackage{tabu}
\usepackage{tabularx}
\usepackage{adjustbox}
\usepackage{array}
\usepackage{booktabs} % For formal tables
\usepackage{wrapfig} % For wraptable environment

\usepackage{multirow}
\usepackage{lipsum}
\usepackage{algorithm}
\usepackage{pifont}
\usepackage{color, soul}
\usepackage{amsmath}
\usepackage[graphicx]{realboxes}
\usepackage{comment}
\usepackage{mdframed}
\usepackage{float}
\usepackage{url}

\usepackage[noend]{algpseudocode}
\usepackage[utf8]{inputenc}
\usepackage{comment}
\usepackage{colortbl}
\usepackage{xcolor}
\usepackage{threeparttable}
\usepackage{subcaption}
\usepackage[acronym]{glossaries}
\usepackage{duckuments}
\usepackage{balance}
\usepackage{environ}
\usepackage{lipsum}
\usepackage[normalem]{ulem}
 \usepackage{adjustbox}
 \usepackage{subcaption}
 \usepackage{rotating}
 \usepackage{makecell}
 \usepackage{graphicx}
\usepackage{booktabs,array,seqsplit}
\usepackage{wrapfig}

\usepackage[most]{tcolorbox}
\usepackage{pgfplots}
\pgfplotsset{compat=1.18}
\usepackage{pgfplotstable}
\usepackage{xcolor}
\usepackage{booktabs}
\usepackage{tikz}
\usetikzlibrary{shapes.geometric,shapes.symbols,positioning,arrows.meta,fit,calc,shapes.misc}

\def\BibTeX{{\rm B\kern-.05em{\sc i\kern-.025em b}\kern-.08em
    T\kern-.1667em\lower.7ex\hbox{E}\kern-.125emX}}

\newcolumntype{P}[1]{>{\centering\arraybackslash}m{#1}}

\definecolor{extra-light-gray}{rgb}{0.95,0.95,0.95}

\NewEnviron{e-box}[1][]{%
    \begin{mdframed}[backgroundcolor=extra-light-gray]
        \BODY
    \end{mdframed}
}

\newcounter{FindingsCounter}
\NewEnviron{finding}[1][]{%
	\refstepcounter{FindingsCounter}
	
    \begin{mdframed}[backgroundcolor=extra-light-gray]
        \textbf{Finding \arabic{FindingsCounter}}: \BODY
    \end{mdframed}
}

\renewcommand{\hl}[1]{\textcolor{orange!80!black}{#1}}

\usepackage{fontawesome5}

\usepackage{listings}
\lstdefinelanguage{ghayaml}{
  morekeywords={name,on,jobs,build,test,deploy,runs-on,steps,uses,run,with,env,
                needs,permissions,timeout-minutes,branches,ref,contents,id-token,
                if,strategy,matrix,container,services,secrets,push,pull_request,
                pull_request_target,schedule,workflow_dispatch,defaults},
  sensitive=true,
  morecomment=[l]{\#},
  morestring=[b]",
  morestring=[b]',
}
\newcommand{\anno}[1]{\textcolor{orange!85!black}{\textbf{\small\textcircled{\scriptsize #1}}}}
\newcommand{\smellmark}{\textcolor{red!70!black}{\faBug}}
\newcommand{\fixmark}{\textcolor{green!55!black}{\faCheckCircle}}
\definecolor{lstwrapgray}{gray}{0.45}
\newcommand{\lstwrap}{\mbox{\textcolor{lstwrapgray}{$\hookrightarrow$} }}
\floatstyle{plaintop}
\newfloat{listing}{tbp}{lol}
\floatname{listing}{Listing}

\definecolor{diffadd}{rgb}{0.2, 0.6, 0.2}
\definecolor{diffremove}{rgb}{0.6, 0.2, 0.2}

\lstdefinelanguage{diff}{
  morecomment=[f][\color{diffadd}]{+},    % Linhas com + ficam verdes
  morecomment=[f][\color{diffremove}]{-}, % Linhas com - ficam vermelhas
  morecomment=[f][\color{gray}]{@@},      % Linhas de metadados ficam cinzas
  morecomment=[f][\color{blue}]{---},     % Cabeçalho do diff
  morecomment=[f][\color{blue}]{+++},     % Cabeçalho do diff
}

\newcommand{\groupcell}[1]{\rotatebox[origin=c]{90}{\textsc{\scriptsize #1}}}

   \usepackage{tikz}
  \usetikzlibrary{arrows.meta, positioning, calc, backgrounds}
   \newcommand{\circled}[1]{\tikz[baseline=(c.base)]\node[circle,draw,
       inner sep=0.6pt,font=\tiny\bfseries](c){#1};\,}
\AtBeginDocument{%
  \providecommand\BibTeX{{%
    Bib\TeX}}}

\setcopyright{none}
\copyrightyear{2027}
\acmYear{2027}
\acmDOI{XXXXXXX.XXXXXXX}

\acmConference[FSE '27]{Proceedings of the 35th ACM Symposium on the Foundations of Software Engineering}{July 12--16, 2027}{Shenzhen, China}
\acmISBN{978-1-4503-XXXX-X/2018/06}

\begin{document}

\title[GHA-SmellFix: Taking Action on Workflow Smells]
{\textsc{GHA-SmellFix}: Taking Action with Agentic LLMs for Detecting and Refactoring GitHub Actions Workflow Smells}


\author{Anderson Uch{\^o}a}
\orcid{0000-0002-6847-5569}
\correspondingauthor
\email{andersonuchoa@ufc.br}
\affiliation{%
  \institution{Federal University of Cear{\'a}}
  \city{Itapaj{\'e}}
  \state{Cear{\'a}}
  \country{Brazil}
}
\affiliation{%
  \institution{North Carolina State University}
  \city{Raleigh}
  \state{North Carolina}
  \country{USA}
}

\author{Jhefferson Souza}
\orcid{0009-0001-2348-2896}
\email{jhefferson.souza@alu.ufc.br}
\affiliation{%
  \institution{Federal University of Cear{\'a}}
  \city{Itapaj{\'e}}
  \state{Cear{\'a}}
  \country{Brazil}
}

\author{Emanuel Ávila}
\orcid{0009-0006-0317-3021}
\email{emanuel.pires@alu.ufc.br}
\affiliation{%
  \institution{Federal University of Cear{\'a}}
  \city{Itapaj{\'e}}
  \state{Cear{\'a}}
  \country{Brazil}
}

\author{Gabriel Silva}
\orcid{0009-0007-2341-1231}
\email{gabrielvsilva@alu.ufc.br}
\affiliation{%
  \institution{Federal University of Cear{\'a}}
  \city{Itapaj{\'e}}
  \state{Cear{\'a}}
  \country{Brazil}
}

\author{Kevin Lira}
\orcid{0000-0001-7324-3925}
\email{kwlira@ncsu.edu}
\affiliation{%
  \institution{North Carolina State University}
  \city{Raleigh}
  \state{North Carolina}
  \country{USA}
}

\author{Hui Sun}
\orcid{0009-0000-7911-3435}
\email{hsun26@ncsu.edu}
\affiliation{%
  \institution{North Carolina State University}
  \city{Raleigh}
  \state{North Carolina}
  \country{USA}
}

\author{Nelson Ara{\'u}jo}
\orcid{0009-0001-5395-5930}
\email{nelsonfelipe@alu.ufc.br}
\affiliation{%
  \institution{Federal University of Cear{\'a}}
  \city{Itapaj{\'e}}
  \state{Cear{\'a}}
  \country{Brazil}
}

\author{Arthur Cordeiro}
\orcid{0009-0009-1090-0532}
\email{arthur.cordeiro32@alu.ufc.br}
\affiliation{%
  \institution{Federal University of Cear{\'a}}
  \city{Itapaj{\'e}}
  \state{Cear{\'a}}
  \country{Brazil}
}

\author{Rohit Gheyi}
\orcid{0000-0002-5562-4449}
\email{Rohit@dsc.ufcg.edu.br}
\affiliation{%
  \institution{Federal University of Campina Grande}
  \city{Campina Grande}
  \state{Para{\'i}ba}
  \country{Brazil}
}

\author{Wesley Assun{\c{c}}{\~a}o}
\orcid{0000-0002-7557-9091}
\email{wguezas@ncsu.edu}
\affiliation{%
  \institution{North Carolina State University}
  \city{Raleigh}
  \state{North Carolina}
  \country{USA}
}

\renewcommand{\shortauthors}{Uchôa et al.}

\begin{abstract}
Continuous integration and continuous delivery workflows have become key artifacts in modern software projects. In GitHub Actions, they encode build, test, release, deployment, and security automation through YAML configuration files. Unfortunately, these workflows often contain recurring configuration smells that can affect security posture, reliability, maintainability, or operational stability.
Existing methods to address GitHub Actions workflow smells have two limitations: (i) they are mostly rule-based; however, rule creation and evolution are difficult tasks, and (ii) they stop at reporting findings; developers must still interpret these findings and craft fixes.
Agentic LLM pipelines are a natural fit for GitHub Actions smell remediation because the task combines contextual detection, transformation selection, and patch generation. Assigning these responsibilities to specialized agents enables the pipeline to condition refactoring on positive detection verdicts and use task-specific reasoning at each stage, potentially avoiding unnecessary modifications. Yet, it remains unclear whether this decomposition improves effectiveness or mainly changes operational overhead.
We present \textsc{GHA-SmellFix}, an agentic LLM pipeline for detecting and refactoring GitHub Actions workflow smells. \textsc{GHA-SmellFix} is instantiated with two orchestration designs: (i) Joint pipeline, a single agent detects and refactors, and (ii) Decomposed pipeline, separate agents detect and then refactor.
To empirically compare these two designs, we operationalize 19 smells from an existing catalog with explicit detection criteria, refactoring expectations, exclusion rules, and safety checks, and build a manually curated ground truth of 1,710 \textit{(workflow, smell)} pairs. We instantiate the two designs with four LLMs (GPT-5.4-nano, Gemini 3.5 Flash, Claude Haiku 4.5, and DeepSeek V4 Flash) and evaluate them on 570 pairs sampled from this ground truth.
Our results show that the key value of agentic decomposition is not universally better patch quality, but better control over whether the agent attempts patch generation. Decomposition particularly helps models prone to missing smells and can reduce operational overhead by gating patch generation. Once a smell is correctly detected, clean refactoring is generally effective, making target-smell detection the primary end-to-end bottleneck. However, false-positive detections can trigger unnecessary edits to execution-relevant workflow constructs. Thus, practical CI/CD agents need not only effective patch generation, but also reliable abstention and construct-aware validation to avoid unnecessary or execution-relevant workflow changes.

\end{abstract}



\maketitle

\section{Introduction}
\label{sec:introduction}

Continuous Integration and Continuous Delivery (CI/CD) pipelines have become key maintenance artifacts in modern software engineering~\cite{hilton2017trade,gallaba2018use,vassallo2019automated,zampetti2020empirical}. GitHub Actions is widely adopted to implement such pipelines through repository-level YAML files, called \textit{workflows}, that automate builds, tests, releases, deployments, and security-related tasks~\cite{cardoen2024dataset,decan2022use}. Although workflows have a declarative syntax~\cite{github2026workflow}, they are not simple configuration artifacts. They may invoke third-party actions, access repository secrets, publish artifacts, define permissions, execute scripts, and interact with deployment environments~\cite{koishybayev2022characterizing}. As a result, the way a workflow is written can affect the security posture~\cite{koishybayev2022characterizing,khatami2024catching}, reliability, maintainability, and operational stability of software projects~\cite{gallaba2018use,vassallo2020configuration,zampetti2020empirical}. Prior studies show that CI/CD workflows frequently contain configuration smells or anti-patterns, i.e., recurring patterns that may increase the risk of vulnerabilities, unstable builds, excessive resource usage, or performance degradation~\cite{khatami2024catching,koishybayev2022characterizing,vassallo2020configuration,zampetti2020empirical,gallaba2018use}.

Existing methods for analyzing quality and security problems in GitHub Actions workflows are largely rule-based~\cite{benedetti2022automatic,freitas2024gash,khatami2024catching}. These methods encode known patterns and provide inexpensive, reproducible screening. However, they have two main limitations. First, their detection logic depends on engineered rules whose design and maintenance can require substantial effort as workflow constructs, platform features, and smell definitions evolve~\cite{decan2022use,mazrae2026empirical,bardsiri2026github}. For instance, checking effective permissions may require combining workflow-level and job-level \texttt{permissions}, while fork-guard smells depend on event triggers, conditional expressions, and repository context rather than on a single YAML key.
Second, only reporting a smell does not complete the maintenance task: currently, developers must still judge whether the finding is relevant, decide how to remediate it, and assess whether the resulting patch is correct or introduces unrelated behavior changes~\cite{johnson2013don,beller2016analyzing,ye2021automated}. This is challenging because workflow refactoring may involve constructs such as permissions, triggers, conditions, runners, dependencies, deployment steps, caches, matrices, and reusable actions or workflows~\cite{bardsiri2026github,mazrae2026empirical}. Thus, the open problem is not only detecting smells, but generating and evaluating candidate refactorings that are structurally valid, scope-bounded, and abstention-aware.

Large Language Models (LLMs) offer a promising mechanism for this broader maintenance task because they can interpret configuration context, reason about intended changes, and generate candidate patches~\cite{hou2024large}. 
However, LLM-generated edits may be risky: a patch may remove the target smell while changing unrelated triggers, permissions, runners, dependencies, or deployment behavior. 
Agentic pipelines can organize LLM capabilities into explicit roles, for instance, by separating smell detection from patch generation~\cite{wang2024survey}. 
Yet, we still lack a clear understanding of how decomposing workflow maintenance tasks affects detection and refactoring performance, as well as its trade-offs in execution latency, token efficiency, and API cost.

This paper presents \textsc{GHA-SmellFix}, an agentic LLM pipeline for identifying and refactoring GitHub Actions configuration smells. Given smell definitions, a workflow file, and repository-level context, \textsc{GHA-SmellFix} detects whether a smell is present and, when appropriate, generates a scope-bounded candidate refactoring. 
We compare two \textsc{GHA-SmellFix} orchestration designs: (D1) a \textit{joint design} that performs detection and, when applicable, refactoring in one pass; and (D2) a \textit{decomposed design} in which a detection agent first decides whether the target smell is present, and a refactoring agent is invoked only for detected smells. These designs differ in task decomposition, prompt context, number of LLM calls, and control flow; therefore, we use them as practical designs for studying effectiveness, patch quality, abstention behavior, and operational overhead, not as an isolated causal test of adding one agent.

To evaluate these designs, we operationalize 19 GitHub Actions smells from an existing catalog~\cite{khatami2024catching}, defining explicit detection criteria, refactoring expectations, exclusion rules, construct-level scope, and risk checks. We then instantiate the two designs using four LLMs (GPT-5.4-nano, Gemini 3.5 Flash, Claude Haiku 4.5, and DeepSeek V4 Flash) and compare them in paired cross-model and cross-design evaluations on 570 annotated \textit{(workflow, smell)} pairs.
These pairs were sampled from a 1,710-pair ground truth manually curated for this study, covering positive and negative cases for 19 smells. To assess generated refactorings, we use a construct-aware evaluation protocol that distinguishes clean smell removals, removals with unexpected edits, removals with potential behavior changes, failed/no-patch cases, and unnecessary edits on ground-truth-negative pairs. We analyze detection effectiveness, end-to-end and conditional refactoring outcomes, abstention behavior, and overhead in terms of latency, token usage, and estimated API cost.

Our results show that \textit{agentic decomposition provides selective rather than universal benefits for GitHub Actions maintenance}. Separating detection from refactoring modestly improves aggregate detection, increasing average F1 from 86.34\% to 87.55\% and recall from 79.12\% to 82.37\%, with the largest gain for GPT-5.4-nano, whose F1 rises from 81.38\% to 87.15\%; models already strong in the joint setting change little. More importantly, the main end-to-end bottleneck is \textit{detecting when the target smell is present rather than generating the patch itself}: clean removal reaches 74\% overall, but rises to 92.6\% once the target smell is correctly detected.
At the same time, successful smell removal does not guarantee patch quality, and abstention remains challenging: the pipelines produced 93 unnecessary edits on negative cases, 54.84\% of which affected behavior-relevant workflow constructs. Decomposition also need not increase operational cost, as the detection gate reduces latency for three of the four evaluated models by avoiding unnecessary refactoring calls. Overall, these findings position agentic LLM pipelines as promising \textit{review-support mechanisms for CI/CD maintenance}, whose practical value depends on model- and smell-aware orchestration, reliable abstention, and construct-aware validation rather than on decomposition alone.

Our contributions include: 
(i) \textsc{GHA-SmellFix}, an agentic LLM pipeline for target-smell detection and candidate refactoring; (ii) a manually curated ground truth of 1,710 annotated \textit{(workflow, smell)} pairs, including positive and negative cases, designed to support paired evaluation of detection and refactoring behavior; 
(iii) a construct-aware evaluation protocol for LLM-generated workflow patches, covering smell removal, workflow validity, expected construct changes, unexpected edits, potential behavior-impacting changes, failed/no-patch cases, and unnecessary edits; and 
(iv) an empirical evaluation of two \textsc{GHA-SmellFix} orchestration designs across four LLMs, covering effectiveness, abstention, construct-level safety, and operational overhead. 

\section{Related Work}
\label{sec:relatedWork}

\textbf{CI/CD anti-patterns and workflow quality.}
Prior work has shown that CI/CD pipelines accumulate recurring configuration problems that affect maintainability, feedback, security, and delivery behavior. Hilton et al.~\cite{hilton2017trade} discuss trade-offs in CI adoption, including assurance, security, and flexibility. Gallaba and McIntosh~\cite{gallaba2018use} show that CI features are often used or misused in ways that affect build behavior and feedback. Zampetti et al.~\cite{zampetti2020empirical} characterize bad practices in CI pipelines, while Vassallo et al.~\cite{vassallo2019automated,vassallo2020configuration} study CI/CD anti-patterns, decay, configuration smells, and linter-based reporting. These studies establish CI/CD configuration quality as a recurring maintenance concern, but they focus mainly on characterizing, detecting, or reporting problematic practices rather than generating and evaluating candidate refactoring.

\noindent\textbf{GitHub Actions workflows, smells, and scanners.}
GitHub Actions brings CI/CD quality concerns to repository-level YAML workflows. Prior work has studied the adoption and evolution of GitHub Actions workflows~\cite{cardoen2024dataset,decan2022use,valenzuela2022evolution,mazrae2026empirical,sun2026test}, their security risks~\cite{koishybayev2022characterizing}, and opportunities for resource-usage optimization~\cite{bouzenia2024resource}. Recent work also shows that workflow reliability and maintenance effort are related to specific constructs, such as triggers, jobs, steps, permissions, matrices, conditions, and reusable components~\cite{bardsiri2026github}. This construct-level perspective is important for refactoring because an edit may be syntactically valid while still changing behavior-relevant parts of the automation.

Several tools and catalogs target GitHub Actions quality and security. Khatami et al.~\cite{khatami2024catching} introduced a catalog of smells and the \textsc{GHA} analyzer. \textsc{GHAST}~\cite{benedetti2022automatic} and \textsc{GASH}~\cite{freitas2024gash} inspect workflows for recurring security or configuration problems, while Fares et al.~\cite{fares2026unpacking} show that workflow security scanners differ substantially in scope, detected weaknesses, and usability. These approaches are useful as scalable detectors, but they stop at reporting findings. They do not assess whether generated patches remove a target smell, stay within the expected construct scope, or abstain when the smell is absent. Our work evaluates generated candidate patches, not only detects smell instances.

\noindent\textbf{LLM-based agents for software maintenance.}
LLM-based agents have recently been applied to software maintenance tasks that involve repository interaction, patch generation, and validation. SWE-agent~\cite{yang2024swe} and OpenHands~\cite{wang2025openhands} frame maintenance as an interactive repository-level task, while RepairAgent~\cite{bouzenia2025repairagent} uses agentic repair loops to plan and validate program fixes. Recent work on test smells shows that agentic language models can support both smell detection and refactoring in source-code-centric artifacts~\cite{melo2025agentic}. GitHub Actions workflows differ from these settings: small YAML edits can alter permissions, event exposure, secret access, runner selection, artifact publication, deployment behavior, or reusable-workflow invocation. Thus, patch quality cannot be assessed only by syntactic validity or smell removal. Moreover, recent evidence cautions against assuming that more complex agentic orchestration necessarily improves software engineering outcomes; simpler localization--repair--validation pipelines can compete with more autonomous agents~\cite{xia2025agentless}. We therefore treat decomposition as an empirical design choice rather than an assumed improvement.

\section{Study Settings}
\label{sec:studySettings}

To design our study, we adopted the Goal-Question-Metric (GQM) template~\cite{caldiera1994goal}. The goal is to \textbf{analyze} \textsc{GHA-SmellFix} orchestration designs; \textbf{for the purpose of} detecting and refactoring configuration smells in GitHub Actions workflows; \textbf{with respect to} (i) target-conditioned detection effectiveness, (ii) refactoring correctness, construct-level consistency, and abstention behavior, and (iii) the effectiveness--overhead trade-offs of task decomposition; \textbf{from the viewpoint of} researchers studying LLM-based CI/CD maintenance; \textbf{in the context of} a manually curated ground truth of \textit{(workflow, smell)} pairs covering 19 configuration smells. The research questions (RQ) are:

\textbf{RQ$_1$:} \textit{How effectively can each \textsc{GHA-SmellFix} orchestration design detect target GitHub Actions configuration smells?} -- This RQ evaluates target-conditioned detection over explicit \textit{(workflow, smell)} pairs. Each prediction is a binary decision about whether the target smell is \textit{present} or \textit{absent}, compared with the annotated ground truth using precision, recall, accuracy, and F1-score.

\textbf{RQ$_2$:} \textit{To what extent can each \textsc{GHA-SmellFix} orchestration design generate structurally valid and scope-bounded refactorings for GitHub Actions workflow smells?} -- This RQ evaluates end-to-end and conditional refactoring outcomes. We examine whether generated patches remove the target smell, preserve YAML and GitHub Actions structural validity, stay within the expected construct scope, and avoid potential behavior-impacting changes. We also analyze unnecessary edits on ground-truth negative pairs, where abstention is expected. 

\textbf{RQ$_3$:} \textit{What is the effectiveness--overhead trade-off of task decomposition in \textsc{GHA-SmellFix}?} -- This RQ compares (D1) a joint detect-and-refactor design with (D2) a decomposed detect-then-refactor design. These designs differ in task decomposition, prompt context, number of LLM calls, and control flow. We compare detection effectiveness, end-to-end and conditional refactoring outcomes, abstention behavior, latency, token usage, and estimated cost. 

\subsection{Study Steps and Procedures}
\label{sec:dataCollection}

\textbf{Step 1: Repository selection and workflow extraction.}
We relied on the GitHub Actions workflow dataset by Cardoen et al.~\cite{cardoen2024dataset} to identify open-source repositories that use GitHub Actions, deriving 49,510 candidate repositories.
We filtered repositories based on criteria adapted from prior studies~\cite{kalliamvakou2014promises,munaiah2017curating,pickerill2020phantom}.
We kept repositories that: 
(i) had at least 100 GitHub stars; (ii) had at least 100 commits; (iii) had activity within the last 12 months; (iv) were at least six months old; (v) used \texttt{main} or \texttt{master} as the default branch; and (vi) contained at least two workflow files. We excluded repositories that were: (i) forks; (ii) archived; (iii) named as learning, tutorial, clone, template, actions, workflows, or reusable examples; or (iv) created primarily for saving workflows or actions. At the workflow level, we kept files that: (i) were triggered by \texttt{pull\_request} or \texttt{push} events; and (ii) had at least 100 execution runs. We excluded workflow files that: (i) had no modification history; (ii) had names or filenames containing documentation-related terms, such as \texttt{docs.yml}; (iii) were empty or contained placeholder content; or (iv) had invalid YAML syntax. With these criteria, we selected a dataset comprising 27,704 repositories and 49,836 unique GitHub Actions workflow files. 
We then used the GitHub API to retrieve repository metadata and collect the current configuration files under the \texttt{.github/workflows/} directory, adopting a snapshot cut-off on April 9, 2026.

\textbf{Step 2: Selection and operationalization of configuration smells.}
We rely on Khatami et al.'s catalog~\cite{khatami2024catching} because it provides fine-grained GitHub Actions smells and tool support through the \textsc{GHA} analyzer, which we use for sampling. The catalog is grounded in evidence of workflow evolution: candidate smells were derived from recurring patterns of workflow change and examined through a contribution study with maintainers.
The catalog contains 22 smells. We selected 19 and excluded \textit{AddCommentsToWorkflows}, \textit{CorrectIndentation}, and \textit{NameRunSteps}, because they mainly capture readability or formatting conventions rather than security, performance, or CI/CD execution concerns. 
We refined the selected smells into explicit detection conditions, exclusion criteria, uncertainty-handling rules, and counting units. These refinements support consistent manual annotation and agent-based evaluation, especially for ambiguous GitHub Actions constructs such as reusable workflows, local actions, matrices, fork guards, and deployment-related steps. We also cross-checked the refined definitions against the official GitHub Actions documentation~\cite{github2026workflow}.

Based on the refined definitions, two authors classified each smell as \textit{localized} or \textit{context-dependent} through discussion and consensus. \textit{Localized smells} rely on explicit evidence within a bounded set of workflow constructs, such as a job, step, action reference, permission block, timeout, or runner declaration. \textit{context-dependent smells} require combining evidence across workflow constructs, scopes, relationships, or repository-level context, such as triggers, fork guards, deployment behavior, concurrency semantics, matrices, reusable workflows, or workflow intent. Table~\ref{tab:smell-catalog} overviews the smells and the core operational criterion for each smell.

\begin{table}[ht!]
\centering
\tiny
\caption{GitHub Actions configuration smells, refined from Khatami et al.~\cite{khatami2024catching}, and ground-truth coverage. P/N denotes positive/negative pairs; full definitions are in the replication package.}
\label{tab:smell-catalog}
\setlength{\tabcolsep}{3.4pt}
\renewcommand{\arraystretch}{1.02}

\begin{tabularx}{\textwidth}{
@{}
>{\centering\arraybackslash}p{0.32cm}
>{\raggedright\arraybackslash}p{2.6cm}
>{\raggedright\arraybackslash}X
>{\centering\arraybackslash}p{0.65cm}
@{}
}
\toprule
\textbf{Type} & \textbf{Configuration smell} & \textbf{Core operational criterion} & \textbf{P/N} \\
\midrule

\multirow{10}{*}{\groupcell{\textbf{Localized}}}
& \texttt{MissingPermExternalActions}
& A job uses an external step-level GitHub action and has no effective workflow- or job-level \texttt{permissions} block.
& 50/50 \\

& \texttt{UnpinnedActionVersion}
& A step-level GitHub action reference uses \texttt{owner/repo@ref}, but \texttt{ref} is not a full 40-character commit SHA.
& 36/36 \\

& \texttt{MissingTokenPermissions}
& A job uses \texttt{GITHUB\_TOKEN} or authenticated GitHub operations without an effective \texttt{permissions} block.
& 50/50 \\

& \texttt{MissingJobTimeout}
& A job defined in the current workflow file does not declare job-level \texttt{timeout-minutes}.
& 48/48 \\

& \texttt{UnsafeArtifactUploadFork}
& A PR-triggered workflow uploads retained GitHub Actions artifacts without an effective fork guard.
& 33/33 \\

& \texttt{MissingUploadCondition}
& A retained artifact-upload step lacks an explicit step-level or job-level \texttt{if} condition.
& 38/38 \\

& \texttt{MutableRunnerVersion}
& A job uses a mutable GitHub-hosted runner label ending in \texttt{-latest}.
& 50/50 \\

& \texttt{SuboptimalCacheUsage}
& A workflow uses \texttt{actions/cache} for dependencies already supported by setup-action built-in caching.
& 49/49 \\

& \texttt{MultiCommandStep}
& A \texttt{run} step combines separable commands from different workflow phases.
& 25/25 \\

& \texttt{UnpinnedDependencies}
& A \texttt{run} step installs external packages or tools without fixed versions or equivalent immutable references.
& 38/38 \\

\midrule

\multirow{9}{*}{
    \rotatebox[origin=c]{90}{
        \shortstack{
            \textsc{\scriptsize \textbf{Context-}}\\
            \textsc{\scriptsize \textbf{dependent}}
        }
    }
}
& \texttt{MissingForkGuardPR}
& An issue/PR-triggered workflow performs issue or pull-request actions without an effective fork guard.
& 50/50 \\

& \texttt{MissingConcurrencyPR}
& A PR-triggered workflow lacks applicable \texttt{concurrency.cancel-in-progress: true} to cancel older runs.
& 50/50 \\

& \texttt{MissingConcurrencyBranch}
& A branch-push workflow lacks applicable \texttt{concurrency.cancel-in-progress: true} to cancel older runs.
& 50/50 \\

& \texttt{MissingPathFilters}
& A CI workflow triggered by \texttt{push} or \texttt{pull\_request} lacks \texttt{paths} or \texttt{paths-ignore} filters.
& 39/39 \\

& \texttt{MissingForkGuardSchedule}
& A scheduled workflow can run outside the intended repository without a repository, owner, or fork-status guard.
& 50/50 \\

& \texttt{UnsafeDeployOnFork}
& A workflow can run in a fork-related PR context and perform external delivery without an effective fork guard.
& 50/50 \\

& \texttt{MissingConcurrencyGlobal}
& A repeated or side-effectful task can overlap because no effective \texttt{concurrency} group is defined.
& 50/50 \\

& \texttt{SingleOSExecution}
& A test job effectively runs on only one operating system and has no multi-OS matrix.
& 50/50 \\

& \texttt{SingleLangVersionCI}
& A CI job validates only one explicit language/runtime version and has no multi-version matrix.
& 49/49 \\

\midrule
\multicolumn{3}{r}{\textbf{Total}} &
\textbf{855/855} \\

\bottomrule
\end{tabularx}
\end{table}

\textbf{Step 3: Ground-truth construction and manual annotation.} Figure~\ref{fig:candidate_flow} summarizes how we built the manually annotated ground truth for the smells in Table~\ref{tab:smell-catalog}.

\definecolor{posdraw}{RGB}{33,102,172}   % positive (blue)
\definecolor{posfill}{RGB}{222,235,247}
\definecolor{negdraw}{RGB}{178,88,24}    % negative (orange)
\definecolor{negfill}{RGB}{253,232,212}
\definecolor{disc}{RGB}{115,115,115}     % discarded (gray)
\definecolor{band}{RGB}{244,244,244}
 
\begin{figure}[!tp]
\centering
\begin{tikzpicture}[
  font=\scriptsize,
  >={Stealth[length=3.5pt,width=3.5pt]},
  base/.style={draw, rounded corners=2pt, align=center, line width=0.6pt,
               inner sep=2pt, minimum height=0.6cm,
               execute at begin node={\hyphenpenalty=10000}},
  neutral/.style={base, fill=white},
  final/.style={base, line width=1pt, fill=white},
  posbox/.style={base, draw=posdraw, fill=posfill},
  negbox/.style={base, draw=negdraw, fill=negfill},
  arr/.style={->, line width=0.65pt},
  line/.style={line width=0.65pt},
  hdr/.style={font=\scriptsize\bfseries, anchor=south},
]
\def\cA{1.47} \def\cB{4.57} \def\cC{7.58} \def\cD{11.6}
\def\bus{9.4} \def\ygt{0.7}
\newcommand{\dropped}[1]{\\[-1pt]{\tiny\color{disc}#1}}
 
% ================= (1) candidate retrieval =================
\node[neutral, text width=2.75cm] (gha) at (\cA,0.45)
  {\textbf{\textsc{GHA}-guided}\\ \textbf{stratified sampling}\\[1pt]
   {\tiny ${\le}\,1$ instance per smell\\ LOC ${\le}\,209$}};
\node[neutral, text width=2.75cm] (src) at (\cA,-0.9)
  {Manually labeled\\ negative workflows of\\ \textit{UnsafeDeployOnFork}};
 
\node[posbox, text width=2.15cm] (flag)   at (\cB, 0.9) {\textsc{GHA}-flagged\\ \textbf{1,416} inspected};
\node[negbox, text width=2.15cm] (unflag) at (\cB, 0.0) {\textsc{GHA}-unflagged\\ \textbf{1,310} inspected};
\node[posbox, text width=2.15cm] (semi)   at (\cB,-0.9) {LLM-mutated\\ \textbf{49} candidates};
 
\draw[arr] (gha.east) -- ++(0.22,0) |- (flag.west);
\draw[arr] (gha.east) -- ++(0.22,0) |- (unflag.west);
\draw[arr] (src) -- (semi);
 
% ================= (2) manual annotation =================
\node[posbox, text width=3.0cm] (p902) at (\cC, 0.9) {\textbf{902} confirmed positives\dropped{514 discarded (absent)}};
\node[negbox, text width=3.0cm] (n944) at (\cC, 0.0) {\textbf{944} confirmed negatives\dropped{366 discarded (present)}};
\node[posbox, text width=3.0cm] (p45)  at (\cC,-0.9) {\textbf{45} validated positives\dropped{4 discarded (invalid)}};
 
\draw[arr] (flag)   -- (p902);
\draw[arr] (unflag) -- (n944);
\draw[arr] (semi)   -- (p45);
 
% ================= (3) balancing and shared subset =================
\node[final, text width=3.4cm] (gt) at (\cD,\ygt)
  {\textbf{Balanced ground truth}\\
   \textcolor{posdraw}{\textbf{855} pos.} + \textcolor{negdraw}{\textbf{855} neg.}\\
   {\tiny incl.\ 45 semi-synthetic pos.}\dropped{undersampled:\\ $-$92 pos., $-$89 neg.}};
\node[final, text width=3.4cm] (shared) at (\cD,-0.86)
  {\textbf{Shared evaluation set}\\
   \textcolor{posdraw}{\textbf{285} pos.} + \textcolor{negdraw}{\textbf{285} neg.}\\[-1pt]
   {\tiny 15 + 15 pairs per smell}};
 
\draw[line] (p902.east) -| (\bus,\ygt);
\draw[line] (p45.east)  -| (\bus,\ygt);
\draw[line] (n944.east) -| (\bus,\ygt);
\fill (\bus,\ygt) circle (1.1pt);
\draw[arr]  (\bus,\ygt) -- (gt.west);
\draw[arr]  (gt.south) -- (shared.north);
 
% ================= stage bands & headers =================
\begin{scope}[on background layer]
  \fill[band, rounded corners=3pt] (-0.12,-1.42) rectangle (5.8,1.38);
  \fill[band, rounded corners=3pt] (5.95,-1.42) rectangle (9.24,1.38);
  \fill[band, rounded corners=3pt] (9.6,-1.42) rectangle (13.6,1.38);
\end{scope}
\node[hdr] at (2.87,1.38) {\circled{1}Candidate retrieval};
\node[hdr] at (\cC, 1.38) {\circled{2}Manual annotation};
\node[hdr] at (\cD, 1.38) {\circled{3}Balancing \& subset};
\end{tikzpicture}
\caption{Overview of ground-truth construction, manual annotation, and shared-set sampling.} %Colors indicate positive, negative, and discarded candidates.
\label{fig:candidate_flow}
\vspace{-0.5cm}
\end{figure}
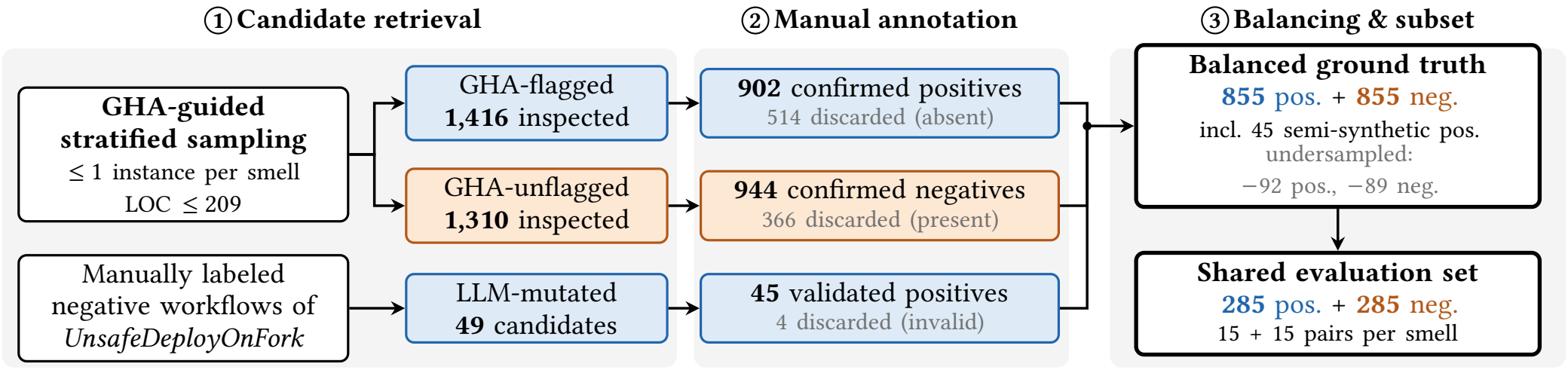

\textbf{\textit{Stratified sampling.}} We adopted pair-based stratified purposive sampling guided, but not labeled, by the \textsc{GHA} analyzer~\cite{khatami2024catching}. For each smell, we sampled \textit{positive} candidates from workflows in which \textsc{GHA} analyzer reported exactly one target-smell instance and \textit{negative} candidates from workflows where the target smell was not flagged. To reduce annotation ambiguity, we kept only workflows with at most one \textsc{GHA}-reported instance of any smell and removed LOC outliers using Tukey's IQR filter~\cite{driscoll1996robustness} (threshold: 209~LOC). We ranked the remaining candidates by ascending \textit{total\_unique\_smells} and descending \textit{workflow\_run\_count}, prioritizing workflows with fewer co-occurring smells and more execution evidence. For each smell, we sampled rounds of 100 positive and 100 negative candidates, repeating for up to six rounds until we reached 50 validated pairs per class, and then balanced each smell by undersampling the larger validated class. The only exception was \textit{UnsafeDeployOnFork}, for which we found only five natural positives. We therefore used Claude Opus 4.8 to mutate real workflows manually labeled negative for this smell, restricting edits to the target-smell construct. After manual validation, we retained 45 of the 49 mutants as positives, which entered balancing together with the natural pairs. We did not synthesize any negative pairs.

\textbf{\textit{Manual annotation.}}
Six annotators worked in three mixed-experience pairs (two master's--undergraduate pairs and one PhD--undergraduate pair). They had one to three years of software engineering experience and prior exposure to GitHub Actions and CI/CD workflows.
Before annotation, we provided annotators with the refined smell catalog, annotation guidelines, and positive/negative examples for each smell. Within each pair, annotators independently inspected the same \textit{(workflow, smell)} pairs, labeled the target smell as \textit{present} or \textit{absent}, and recorded the number of instances. We replaced pairs in which annotators identified multiple target-smell instances using the same iterative procedure, and each team resolved disagreements by consensus. We measured pre-consensus agreement over the 1,710 evaluated \textit{(workflow, smell)} pairs from 902 unique workflow files using Cohen's $\kappa$~\cite{mchugh2012interrater}, obtaining almost perfect agreement~\cite{landis1977measurement} ($\kappa = 0.848$); we report team-level values in the replication package.
We discarded candidates whose manual label contradicted their sampling stratum rather than relabeling them into the natural ground truth (Figure~\ref{fig:candidate_flow}). After adding the 45 validated semi-synthetic positives and balancing per smell, the final ground truth comprised 855 positive and 855 negative pairs, with workflows ranging from 7 to 207 LOC.
From this ground truth, we sampled a shared cross-model evaluation set of 15 positive and 15 negative pairs per smell ($n=570$). Table~\ref{tab:smell-catalog} also summarizes their distribution; per-smell workflow-size statistics and pair-level agreement results are available in our replication package.

\begin{figure}[!tp]
\centering
\resizebox{\columnwidth}{!}{%
\begin{tikzpicture}[
agent/.style={
draw, rounded corners=3pt, thick,
minimum width=2.45cm, minimum height=0.95cm,
align=center, fill=blue!10, font=\scriptsize\sffamily
},
io/.style={
circle, draw, thick,
minimum size=0.95cm, align=center,
fill=orange!15, font=\scriptsize\sffamily
},
arr/.style={->, thick},
lbl/.style={font=\scriptsize\sffamily, fill=white, inner sep=1pt, align=center},
subtitle/.style={font=\scriptsize\sffamily\bfseries}
]

% C1 -- Joint design
\node[io] (c1in) at (0,0) {In};
\node[agent] (c1a) at (2.8,0) {\shortstack{Detection and\\Refactoring}};
\node[io] (c1out) at (5.6,0) {Out};

\draw[arr] (c1in) -- (c1a);
\node[lbl] at (1.1,1) {\shortstack{workflow, target smell,\\repository context}};

\draw[arr] (c1a) -- (c1out);
\node[lbl] at (4.4,1) {\shortstack{verdict and optional\\candidate patch}};

\node[subtitle] at (2.8,1.75) {D1 -- Joint Design};

% C2 -- Decomposed design
\node[io] (c2in) at (7.7,0) {In};
\node[agent] (c2d) at (10.4,0) {Detection};
\node[agent] (c2r) at (13.6,0) {Refactoring};
\node[io] (c2out) at (16.4,0) {Out};

\draw[arr] (c2in) -- (c2d);
\node[lbl] at (8.7,1) {\shortstack{workflow, target smell,\\repository context}};

\draw[arr] (c2d) -- (c2r);
\node[lbl] at (12.0,0.79) {\shortstack{present and\\detection result}};

\draw[arr] (c2r) -- (c2out);
\node[lbl] at (15.5,0.58) {\shortstack{candidate\\patch}};

\draw[arr] (c2d.north) -- (10.4,1.3) -- (16.4,1.3) -- (c2out.north);
\node[lbl] at (13.4,1.33) {\shortstack{absent: original workflow returned}};

\node[subtitle] at (12.0,1.75) {D2 -- Decomposed Design};

\end{tikzpicture}
}
\vspace{-5mm}
\caption{\textsc{GHA-SmellFix} orchestration designs for target-conditioned smell detection and refactoring.}
\label{fig:agentic_configurations}
\end{figure}
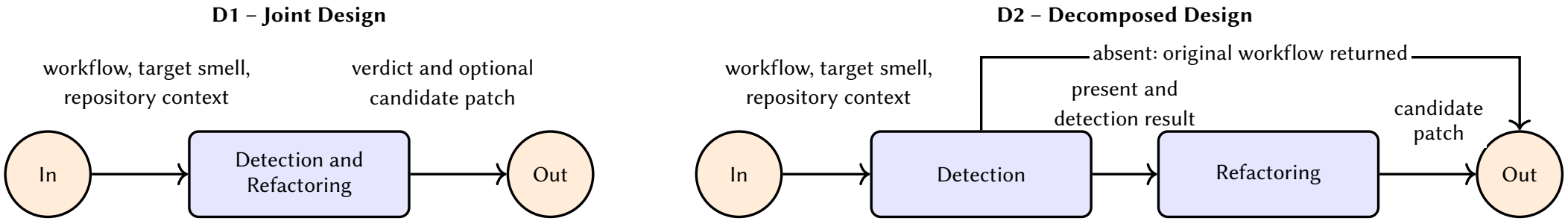

\textbf{Step 4: \textsc{GHA-SmellFix} orchestration designs.}
We instantiate \textsc{GHA-SmellFix} with two agentic LLM orchestration designs that differ in task decomposition and control flow, while keeping the input format, smell catalog, output schemas, and underlying LLM fixed within each comparison.
Figure~\ref{fig:agentic_configurations} illustrates the two designs.
D1 is a \textit{joint detect-and-refactor design}: a single agent classifies the target smell as ``present'' or ``absent'' and, when applicable, returns a candidate patch. D2 is a \textit{decomposed detect-then-refactor design}: a detection agent first classifies the target smell; if the smell is predicted as ``absent'', the original workflow is returned unchanged; otherwise, a refactoring agent generates a candidate patch. 
The agents are not given repository-level CLI access, external validators, or runtime services, allowing us to assess basic agentic capability. 

We implemented both designs using the \textsc{Agno} framework~\cite{derouiche2025agentic}, with XML-tagged prompts and Pydantic schemas for structured outputs. Prompts combined role-based prompting with reasoning-oriented instructions and prompt-level constraints~\cite{schulhoff2024prompt}. These guardrails constrained agents to the target smell, required the smallest scope-bounded edit, discouraged unrelated workflow changes, enforced raw YAML outputs, and required confidence justifications grounded in workflow evidence.
We logged execution metadata for each LLM call, including token usage and latency, to support overhead analysis. Complete prompt templates and schemas are provided in the replication package.

Our study focuses on target-conditioned detection. GHA-SmellFix receives the target smell definition from our refined catalog, the workflow, and repository-level context (owner/repo, owner, default branch, and file path) to interpret repository-specific locations. We executed D1 and D2 on the shared 570-pair set for paired comparison; the full 1,710-pair ground truth serves as complementary evidence for DeepSeek V4 Flash.

\textbf{Step 5: LLM selection and execution.} 
We selected four LLMs, presented in Table~\ref{tab:evaluated_llms}, varying in provider, family, scale, and access mode. We prioritized code-related capabilities, instruction following, and support for structured outputs because LLM effectiveness on SE tasks varies across models and deployment settings~\cite{jiang2026survey,hou2024large,patcas2026evaluation}, and agentic pipelines depend on role-specific instructions and structured responses~\cite{liu2024agentbench,yang2024swe,bouzenia2025repairagent,shinn2023reflexion}.
We kept the input artifacts, role-specific prompt templates, and structured output schemas fixed across agents, changing only the underlying LLM. 
Between D1 and D2, the workflow input and target smell also remained fixed, while the orchestration changed by decomposing the task, changing prompt context, and adding a detection gate before refactoring.
We used temperature 0 whenever supported to reduce sampling variability. GPT-5.4-nano and Claude Haiku 4.5 required temperature 1 in their thinking modes and were executed $k=3$ times to assess detection stability. To control execution cost, GPT-5.4-nano, Claude Haiku 4.5, and Gemini 3.5 Flash were evaluated on a balanced subset with 15 positive and 15 negative pairs per smell. Gemini 3.5 Flash was executed once.
DeepSeek V4 Flash was evaluated once on the 1,710-pair ground truth. Overall, we ran $(2 \times 2 \times 570 \times 3) + (2 \times 1 \times 570) + (2 \times 1 \times 1{,}710)=11{,}400$ pipeline runs.

\begin{table}[ht!]
\centering
\scriptsize
\caption{Overview of the selected LLMs. Costs are reported in USD per 1M tokens (June 2026).}
\label{tab:evaluated_llms}
\begin{adjustbox}{max width=\columnwidth}
\begin{tabular}{l l l r c c r r}
\toprule
\textbf{Model}
& \textbf{Access}
& \textbf{Type}
& \textbf{Params.}
& \textbf{Temp.}
& \textbf{Thinking}
& \textbf{In (1M/USD)}
& \textbf{Out (1M/USD)} \\
\midrule
GPT-5.4-nano
& OpenAI API
& Proprietary
& --
& 1
& Medium
& 0.20
& 1.25 \\

Claude Haiku 4.5
& Anthropic API
& Proprietary
& --
& 1
& 2,048 tokens
& 1.00
& 5.00 \\

Gemini 3.5 Flash
& Gemini API
& Proprietary
& --
& 0
& Medium
& 1.50
& 9.00 \\

DeepSeek V4 Flash
& OpenRouter API
& Open-weight
& 284B (13B active)
& 0
& 2,048 tokens
& 0.064
& 0.251 \\
\bottomrule
\end{tabular}
\end{adjustbox}
\begin{minipage}{\columnwidth}
\centering
\tiny %For proprietary models, parameter counts are not publicly disclosed. 
GPT-5.4-nano and Gemini 3.5 Flash expose reasoning effort as a qualitative level rather than an explicit token budget; their token allocation is provider-managed.
\end{minipage}
\end{table}

\textbf{Step 6: Data Analysis.} We based our analysis on the shared set of 570 ground-truth \textit{(workflow, smell)} pairs, covering 19 smells with 15 positive and 15 negative instances each.  
The full ground-truth results for DeepSeek V4 Flash are provided in the replication package.

\textbf{\textit{Detection Analysis (RQ$_1$).}}
We assess detection effectiveness using Precision, Recall, F1-score, and Accuracy~\cite{fawcett2006introduction}. Each prediction is a binary decision indicating whether a target smell is present in a workflow;
Accuracy therefore captures the proportion of correct present/absent classifications over positive and negative \textit{(workflow, smell)} pairs. We also assess detection stability for GPT-5.4-nano and Claude Haiku 4.5, which were executed three times under the same setup. Following prior work~\cite{gheyi2026foundation}, we report \textit{accuracy spread}, the difference between maximum and minimum accuracy across the $k=3$ executions for each (model, design) pair. This stability analysis is restricted to detection and to models executed three times.

\textbf{\textit{Refactoring Assessment (RQ$_2$).}} 
Figure~\ref{fig:rq2_assessment} summarizes our assessment protocol. We evaluate patches along three dimensions: \textit{smell removal}, \textit{workflow validity}, and \textit{construct-level consistency}. Because refactoring depends on detection, we report \textit{end-to-end} outcomes over all ground-truth-positive outputs and \textit{conditional} outcomes restricted to correctly detected positives.
For \textit{workflow validity}, we check YAML parsability and required GitHub Actions structure; invalid outputs are unsuccessful refactorings. For the other dimensions, we diff original and refactored workflows, extract modified YAML paths, and normalize them into GitHub Actions constructs using Bardsiri et al.~\cite{bardsiri2026github}. Of their 197 constructs, we use the 117 relevant to our 19 smells and map each smell to its expected modification scope. For instance, \textit{MissingJobTimeout} targets \textit{jobs.\textless job\_id\textgreater.timeout-minutes}, whereas \textit{SingleOSExecution} may involve \textit{jobs.\textless job\_id\textgreater.strategy.matrix.\textless matrix\_var\textgreater}. These mappings define expected fix scope rather than fixed templates and are provided in the replication package.

\providecolor{posdraw}{RGB}{33,102,172}  \providecolor{posfill}{RGB}{222,235,247}
\providecolor{negdraw}{RGB}{214,110,40}  \providecolor{negfill}{RGB}{253,228,208}
\providecolor{stepblue}{RGB}{33,110,190}
\providecolor{band}{RGB}{238,239,242}
\providecolor{panel}{RGB}{247,248,250}
 
\tikzset{person/.pic={\fill (0,0.1) circle (0.045);
  \fill (-0.075,-0.05) .. controls (-0.075,0.045) and (0.075,0.045) .. (0.075,-0.05) -- cycle;}}
 
\begin{figure}[!tp]
\centering
\begin{tikzpicture}[
  font=\scriptsize,
  >={Stealth[length=4pt,width=4pt]},
  nohyph/.style={execute at begin node={\hyphenpenalty=10000\exhyphenpenalty=10000}},
  base/.style={draw, rounded corners=3pt, align=center, line width=0.8pt, inner sep=2.5pt, nohyph},
  chk/.style={base, draw=black!45, fill=white, text width=2.7cm, anchor=north},
  pill/.style={rounded corners=2pt, align=center, inner sep=1.3pt, minimum height=0.32cm,
               text width=2.6cm, font=\tiny, nohyph, anchor=north},
  pp/.style={pill, fill=posfill},
  pn/.style={pill, fill=negfill!80},
  gt/.style={font=\scriptsize\bfseries, anchor=north, inner sep=1pt},
  rep/.style={base, draw=black, fill=white, line width=1pt, text width=1.78cm, minimum height=0.62cm},
  arr/.style={->, line width=0.7pt, shorten >=2pt, shorten <=1.5pt},
  ln/.style={line width=0.7pt},
  byp/.style={->, line width=0.7pt, dashed, draw=black!55, rounded corners=4pt, shorten >=2pt, shorten <=1.5pt},
  note/.style={font=\scriptsize\bfseries, text=black!60, inner sep=1pt, fill=panel},
  num/.style={circle, fill=stepblue, text=white, font=\footnotesize\bfseries, inner sep=0pt, minimum size=0.46cm},
  hdr/.style={font=\scriptsize\bfseries, anchor=west},
  sub/.style={font=\tiny, text=black!70},
]
 
\def\ytop{2.28}

% ======================= bands, panels, headers =======================
% (bands are drawn at the end, sized to the content)
 
\foreach \x/\n/\t in {
  0.26/1/Input,
  1.95/2/Automated processing,
  5.52/3/Blinded assessment,
  11.42/4/Reported views}{
  \node[num] at (\x,3.03) {\n};
  \node[hdr] at (\x+0.26,3.03) {\t};
}
 
% RQ2 dimensions
\node[font=\tiny, align=center, text=black!70, inner sep=1pt] at (8.16,2.64)
  {\textit{Dimensions: smell removal $\cdot$ validity $\cdot$ construct-level consistency}};
 
% ======================= (1) input =======================
\node[base, draw=black!45, fill=white, text width=1.24cm, minimum height=2.5cm] (in) at (0.79,0.72)
  {\textbf{\footnotesize 4,560}\\[-1pt] \textbf{outputs}\\[2pt]
   {\tiny 4 models}\\[-2pt] {\tiny $\times$ 2 designs}\\[-2pt] {\tiny $\times$ 570 pairs}\\[3pt]
   {\color{posdraw}\textbf{2,280}}\\[-2pt] {\tiny\color{posdraw}GT-positive}\\[1pt]
   {\color{negdraw}\textbf{2,280}}\\[-2pt] {\tiny\color{negdraw}GT-negative}};
 
% ======================= (2) automated processing =======================
\node[chk] (val) at (3.4,2.02) 
  {\textbf{Workflow validity}\\[-1pt] {\tiny\color{black!70}YAML parsing + GHA structure}};
 
\node[chk] (dif) at ($(val.south)+(0,-0.3)$) 
  {\textbf{Diff vs.\ original}\\[-1pt] {\tiny\color{black!70}per output}};
 
\node[chk] (con) at ($(dif.south)+(0,-0.3)$) 
  {\textbf{Construct extraction}\\[-1pt]
  {\tiny\color{black!70}117/197 relevant constructs~\cite{bardsiri2026github}}};
 
\draw[arr] (val) -- (dif);
\draw[arr] (dif) -- (con);
\draw[arr] (in.east |- val) -- (val);
 
\def\xb{5.08}
\foreach \n in {val,dif,con} 
  \draw[ln] (\n.east) -- (\xb,0 |- \n.east);
 
% ======================= (3) blinded manual assessment =======================
\def\xm{8.16} 
\def\dx{1.38} 
\def\hw{2.78}
 
\node[gt, anchor=north west] (pt) at (\xm-\hw+0.1,2.0) 
  {Positive pairs {\mdseries\tiny\color{black!65}(smell removal + scope)}};
 
\node[pp] (p1) at ($(\xm,0 |- pt.south)+(-\dx,-0.05)$) {Clean Removal};
\node[pp] (p4) at ($(\xm,0 |- pt.south)+(\dx,-0.05)$) {Failed or No Patch};
\node[pp, minimum height=0.66cm] (p2) at ($(p1.south)+(0,-0.05)$) {Removal w/\\ Unexpected Changes};
\node[pp, minimum height=0.66cm] (p3) at ($(p4.south)+(0,-0.05)$) {Potential\\ Behavior-Impacting Change};
 
\path let \p1=(p2.south), \p2=(p3.south) in coordinate (pbot) at (\xm,{min(\y1,\y2)});
\coordinate (pbot) at ($(pbot)+(0,-0.08)$);
\draw[draw=posdraw, line width=1.1pt, rounded corners=4pt] 
  ($(\xm-\hw,0 |- pt.north)+(0,0.03)$) rectangle (\xm+\hw,0 |- pbot);
 
\node[gt, anchor=north west] (nt) at ($(\xm-\hw+0.1,0 |- pbot)+(0,-0.26)$) 
  {Negative pairs {\mdseries\tiny\color{black!65}(modified outputs only)}};
 
\node[pn, minimum height=0.5cm] (n1) at ($(\xm,0 |- nt.south)+(-\dx,-0.05)$) {Cosmetic Edit};
\node[pn, minimum height=0.5cm] (n2) at ($(\xm,0 |- nt.south)+(\dx,-0.05)$) {Localized\\ Unnecessary Edit};
\node[pn, text width=5.36cm] (n3) at ($(\xm,0 |- n1.south)+(0,-0.05)$) {Behavior-Relevant Unnecessary Edit};
 
\coordinate (nbot) at ($(n3.south)+(0,-0.08)$);
\draw[draw=negdraw, line width=1.1pt, rounded corners=4pt] 
  ($(\xm-\hw,0 |- nt.north)+(0,0.03)$) rectangle (\xm+\hw,0 |- nbot);
 
\coordinate (pmid) at ($(\xm-\hw,0 |- pt.north)!0.5!(\xm-\hw,0 |- pbot)$);
\coordinate (nmid) at ($(\xm-\hw,0 |- nt.north)!0.5!(\xm-\hw,0 |- nbot)$);
 
\draw[arr, shorten <=0pt] (\xb,0 |- pmid) -- (pmid);
\draw[arr, shorten <=0pt] (\xb,0 |- nmid) -- (nmid);
\draw[ln] (\xb,0 |- val.east) -- (\xb,0 |- con.east);
 
% ======================= (4) reported views =======================
\coordinate (pE) at (\xm+\hw,0 |- pmid);
\coordinate (fk) at (11.22,0 |- pmid);
 
\node[rep] (e2e) at (12.63,0.2 |- p1.north) 
  {\textbf{End-to-end}\\[-1.5pt] {\tiny all GT positives}};
 
\node[rep, anchor=north] (cnd) at ($(e2e.south)+(0,-0.12)$) 
  {\textbf{Conditional}\\[-1.5pt] {\tiny GT positives}\\[-2pt] {\tiny verdict = \textit{present}}};
 
\node[rep] (abs) at (12.63,0 |- nmid) 
  {\textbf{Abstention}\\[-1.5pt] {\tiny all GT negatives}};
 
\draw[ln, shorten <=1.5pt] (pE) -- (fk);
\draw[ln] (fk |- e2e) -- (fk |- cnd);
\fill (fk) circle (1.3pt);
\draw[arr, shorten <=0pt] (fk |- e2e) -- (e2e.west);
\draw[arr, shorten <=0pt] (fk |- cnd) -- (cnd.west);
\draw[arr] (\xm+\hw,0 |- nmid) -- (abs.west);
 
% ======================= automatic assignments =======================
\coordinate (ptop) at ($(\xm-\hw,0 |- pt.north)+(0,0.03)$);
\draw[byp] (val.north) -- (val.north |- 0,2.34) -| (p4.north |- ptop);
\node[note, font=\tiny\bfseries] at (7.0,2.34) {invalid $\Rightarrow$ Failed or No Patch};
 
% bottom channel placed below the lowest box, whatever the font metrics
\coordinate (lowest) at ($(con.south)+(0,0)$);
\path let \p1=(con.south), \p2=(nbot), \p3=(in.south) in
  coordinate (lowest) at (0,{min(\y1,\y2,\y3)});
\coordinate (ych) at ($(lowest)+(0,-0.24)$);
\draw[byp] (con.south) -- (con.south |- ych) -- (abs.south |- ych) -- (abs.south);
\node[note, font=\tiny\bfseries] at (\xm,0 |- ych) {unmodified $\Rightarrow$ Correct Abstention};
\coordinate (ypb) at ($(ych)+(0,-0.18)$);
 
% ======================= bands, panels (sized to content) =======================
\begin{scope}[on background layer]
  \foreach \xa/\xb in {0/1.58, 1.68/5.14, 5.24/11.08, 11.18/13.7}{
    \fill[band, rounded corners=4pt] ($(\xa,0 |- ypb)+(0,-0.08)$) rectangle (\xb,3.28);
    \fill[panel, rounded corners=3pt] ($(\xa+0.08,0 |- ypb)$) rectangle (\xb-0.08,2.78);
  }
\end{scope}
 
% ======================= human validation strip =======================
\coordinate (sc) at ($(0,0 |- ypb)+(0,-0.58)$);   % strip centre line
\draw[fill=band, draw=black!25, rounded corners=4pt, line width=0.6pt]
  ($(sc)+(0,0.4)$) rectangle ($(sc)+(13.7,-0.4)$);
 
\pic[scale=2.2, black!75] at ($(sc)+(0.36,-0.105)$) {person};
\pic[scale=2.5, black!75] at ($(sc)+(0.62,-0.12)$) {person};
\pic[scale=2.2, black!75] at ($(sc)+(0.88,-0.105)$) {person};
 
\node[anchor=west, align=left, text width=5.7cm, font=\tiny, nohyph] at ($(sc)+(1.15,0)$)
  {Two blinded annotators label each output independently (smell definition, expected construct scope, rationale, diff, and extracted constructs).};
 
\draw[black!30, line width=0.6pt] ($(sc)+(7.08,0.24)$) -- ($(sc)+(7.08,-0.24)$);
 
\pic[scale=2.5, black!75] at ($(sc)+(7.38,-0.12)$) {person};
 
\node[anchor=west, align=left, text width=5.8cm, font=\tiny, nohyph] at ($(sc)+(7.62,0)$)
  {A third author adjudicates disagreements; both \textit{uncertain} $\Rightarrow$ not assessable. Runtime-dependent facts are not validated. Pre-consensus $\kappa = 0.758$.};
 
\end{tikzpicture}
\caption{Overview of the refactoring assessment protocol. Dashed arrows indicate automatic assignments for invalid positive outputs and unmodified negative outputs; all other outputs are assessed by blinded annotators. Conditional rates use GT-positive outputs whose structured detection verdict is \textit{present}.}
\label{fig:rq2_assessment}
\vspace{-0.5cm}
\end{figure}
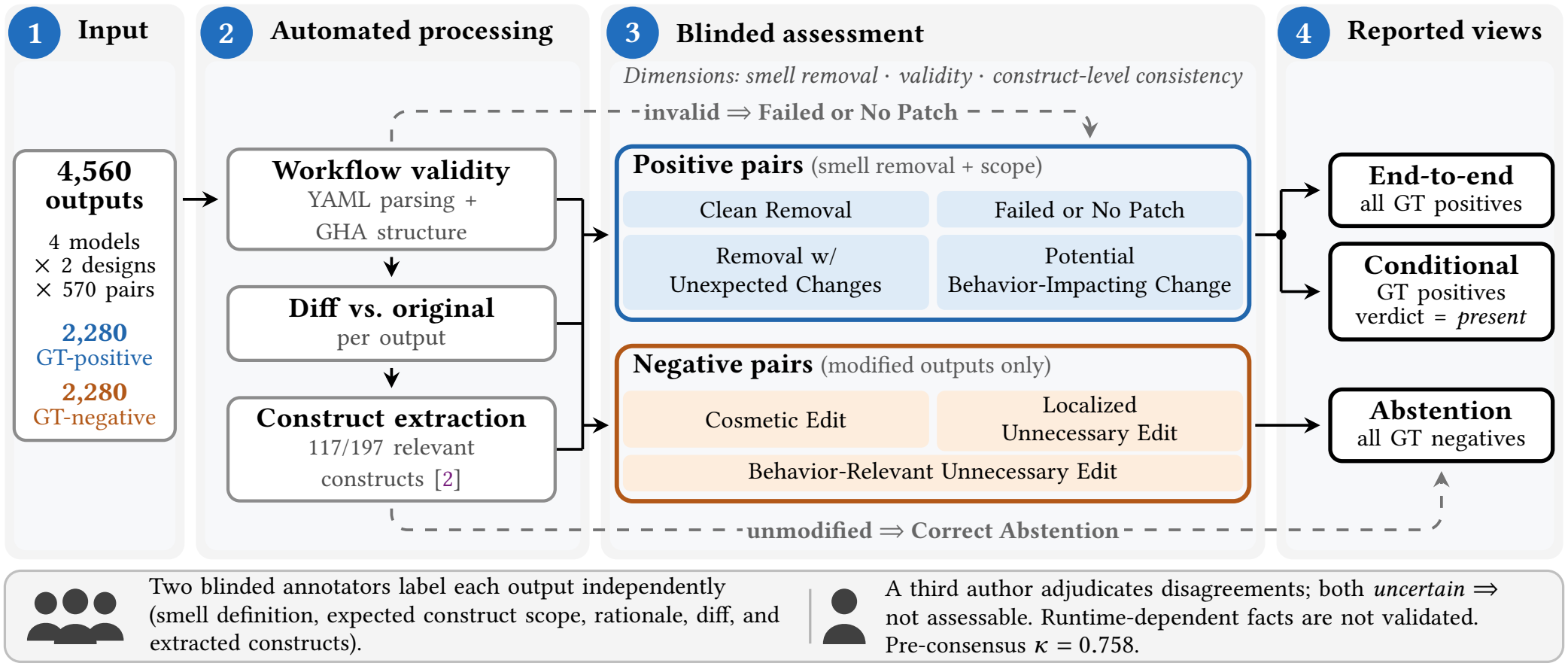
 
We manually inspected 4,560 outputs: 2,280 positive-pair outputs for refactoring assessment and 2,280 negative-pair outputs for abstention. The ground-truth annotator pairs assessed outputs blindly to model and design using the smell definition, expected construct scope, agent rationale, diff, and modified constructs. Because multiple valid fixes may exist, no single reference patch was used.
Positive outputs were classified as \textit{Clean Removal} when the smell was removed, validity preserved, and only expected constructs changed; \textit{Removal with Unexpected Changes} when unrelated constructs were modified; \textit{Removal with Potential Behavior-Impacting Change} when changes could affect workflow execution, security posture, or deployment behavior; or \textit{Failed or No Patch} when the
smell was not removed, validity checks failed, or no applicable patch was produced.

For negative pairs, where abstention was expected, modified outputs were classified as \textit{Cosmetic Edit} limited to formatting, comments, or naming; \textit{Localized Unnecessary Edit}, for plausible but unnecessary smell-related changes; or \textit{Behavior-Relevant Unnecessary Edit}, for changes to execution-relevant constructs. If multiple edit types occurred, we assigned the most behavior-relevant category.
Annotators could mark cases as \textit{uncertain} when static evidence was insufficient. Disagreements were adjudicated by a third author, except when both annotators independently marked a case uncertain; these were treated as not assessable. Pre-consensus agreement was substantial ($\kappa = 0.758$)~\cite{landis1977measurement,mchugh2012interrater}. We do not validate runtime-dependent facts such as commit hashes, package versions, credentials, external actions, or services.

\textbf{\textit{Orchestration effect (RQ$_3$).}}
We compare D1 and D2 as practical \textsc{GHA-SmellFix} designs to assess whether task decomposition improves effectiveness or mainly changes overhead. This comparison does not isolate the effect of adding one agent. We reuse RQ$_1$ metrics for detection, RQ$_2$ outcomes for refactoring, and compare latency, token usage, and estimated API cost for overhead.
For each LLM, we compare D1 and D2 outputs over the same \textit{(workflow, smell)} pairs using McNemar's exact test~\cite{mcnemar1947note} for detection correctness and end-to-end clean removal, and paired Wilcoxon signed-rank tests~\cite{wilcoxon1945individual} for latency and estimated cost, with Holm correction~\cite{holm1979simple} within each metric.

\textbf{\textit{Execution protocol and stability.}}
For GPT-5.4-nano and Claude Haiku 4.5, which were non-deterministic in our setup, we ran each model--design pair three times under D1 and D2. We use the first run in the main analyses to avoid post-hoc run selection, and use the repetitions only for stability analysis. The maximum accuracy spread is 2.11\%, indicating low run-to-run variation. Gemini 3.5 Flash and DeepSeek V4 Flash were configured with temperature zero and executed once; therefore, stability conclusions are limited to GPT-5.4-nano and Claude Haiku 4.5. All repeated-run outputs are provided in the replication package.

\section{Results and Discussion}
\label{sec:results}

\subsection{Detection Effectiveness (RQ$_1$)}
\label{sec:rq1}

To answer \textbf{RQ$_1$},
we analyze detection from two perspectives: (i) \textit{aggregate effectiveness}, comparing models and designs to assess whether task decomposition changes detection; and (ii) \textit{smell-level effectiveness}, examining how these effects vary across the 19 smells and across models.

\begingroup
\setlength{\columnsep}{6pt}   % espaço horizontal entre texto e tabela
\setlength{\intextsep}{4pt}  
\begin{wraptable}{r}{7.5cm}
\centering
\tiny
%\caption{Overall detection effectiveness on the 570 ground-truth (workflow, smell) pairs.}
\caption{Overall detection effectiveness across designs. The best and second-best model values per configuration are shown in \textbf{bold} and \underline{underlined}, respectively.}
\label{tab:rq1_overall_detection}
\begin{adjustbox}{max width=\linewidth}
\begin{tabular}{llrrrrr}
\toprule
\textbf{Design}
& \textbf{Model}
& \textbf{$n$}
& \textbf{F1$\uparrow$}
& \textbf{Pr$\uparrow$}
& \textbf{Re$\uparrow$}
& \textbf{Acc$\uparrow$} \\
\midrule
\multirow{5}{*}{\textbf{D1}}
& Claude Haiku 4.5   & 570  & \textbf{88.81\%} & 93.08\% & \textbf{84.91\%} & \textbf{89.30\%} \\
& DeepSeek V4 Flash  & 570  & \underline{88.27\%} & 94.05\% & \underline{83.16\%} & \underline{88.95\%} \\
& Gemini 3.5 Flash   & 570  & 86.89\% & \textbf{98.23\%} & 77.89\% & 88.25\% \\
& GPT-5.4-nano       & 570  & 81.38\% & \underline{96.17\%} & 70.53\% & 83.86\% \\
& \textbf{Avg.}       & --   & \textbf{86.34\%} & \textbf{95.38\%} & \textbf{79.12\%} & \textbf{87.59\%} \\
\midrule
\multirow{5}{*}{\textbf{D2}}
& Claude Haiku 4.5   & 570  & \textbf{88.52\%} & \underline{93.73\%} & \textbf{83.86\%} & \textbf{89.12\%} \\
& DeepSeek V4 Flash  & 570  & \underline{87.99\%} & 92.97\% & \underline{83.51\%} & \underline{88.60\%} \\
& GPT-5.4-nano       & 570  & 87.15\% & 92.86\% & 82.11\% & 87.89\% \\
& Gemini 3.5 Flash   & 570  & 86.53\% & \textbf{94.21\%} & 80.00\% & 87.54\% \\
& \textbf{Avg.}       & --   & \textbf{87.55\%} & \textbf{93.44\%} & \textbf{82.37\%} & \textbf{88.29\%} \\
\bottomrule
\end{tabular}
\end{adjustbox}
\end{wraptable} 
%\end{table}

\textbf{Task decomposition modestly improves aggregate detection effectiveness.} Table~\ref{tab:rq1_overall_detection} overviews the aggregate detection results.
We observe that D2 achieves the highest aggregate detection effectiveness, reaching an average F1-score of 87.55\% and accuracy of 88.29\%. This gain is mainly recall-driven: recall increases from 79.12\% in D1 to 82.37\% in D2, while precision remains high but decreases from 95.38\% to 93.44\%. Thus, task decomposition shifts the precision--recall balance toward recovering more true smell instances, at the cost of a small increase in false positives. However, this effect is model-dependent. GPT-5.4-nano benefits the most from D2, improving from 81.38\% to 87.15\% F1, driven by a recall increase from 70.53\% to 82.11\%. Claude Haiku 4.5 and DeepSeek V4 Flash are already strong under D1 and remain stable across designs, while Gemini 3.5 Flash shifts from high precision in D1 to higher recall in D2.

The results also suggest that orchestration effects are not explained by model availability alone. DeepSeek V4 Flash, the open-weight model in our study, achieves the second-best F1-score in both designs, closely matching Claude Haiku 4.5, but changes little under decomposition. In contrast, GPT-5.4-nano benefits substantially from D2. Overall, D2 improves aggregate detection, but not uniformly across models; its main benefit is increasing recall, especially for models that under-detect in the joint detect-and-refactor design.

\begin{finding}
Task decomposition modestly improves aggregate detection, raising average F1 from 86.34\% to 87.55\% and recall from 79.12\% to 82.37\%.
\end{finding}

\textbf{Task decomposition helps selectively, depending on smell semantics and model behavior.} Table~\ref{tab:rq1_smell_level_detection_effectiveness} shows that D2 gains vary by smell and model. Overall, we observe high precision across configurations, while recall varies more substantially. Smells with explicit workflow evidence are easier to detect: \textit{MissingPermExternalActions} and \textit{MissingTokenPermissions} reach 100\% F1 in nearly all (model, design) pairs, and \textit{MissingPathFilters}, \textit{UnsafeArtifactUploadFork}, and \textit{SuboptimalCacheUsage} remain close to 100\% in most cases. For these construct-explicit smells, decomposition provides little additional value because the single-agent configuration already performs well.

\begin{table*}[ht!]
\centering
\tiny
\caption{Smell-level detection effectiveness by model and \textsc{GHA-SmellFix} orchestration design. Best F1-score per smell/model is in \textbf{bold}; second-best distinct F1-score is \underline{underlined}. The Avg. row reports macro-averages across smells for each model/design pair. Gray rows highlight selected difficult smells discussed in the text.}

\label{tab:rq1_smell_level_detection_effectiveness}
\setlength{\tabcolsep}{1.0pt}
\renewcommand{\arraystretch}{1.08}

\begin{adjustbox}{max width=\textwidth}
\begin{tabular}{
@{}
>{\centering\arraybackslash}p{0.65cm}
>{\raggedright\arraybackslash}p{3.0cm}
!{\vrule width 0.35pt}
rrrr!{\vrule width 0.25pt}rrrr
!{\vrule width 0.35pt}
rrrr!{\vrule width 0.25pt}rrrr
!{\vrule width 0.35pt}
rrrr!{\vrule width 0.25pt}rrrr
!{\vrule width 0.35pt}
rrrr!{\vrule width 0.25pt}rrrr
@{}
}
\toprule

\multirow{3}{*}{\textbf{Type}}
& \multirow{3}{*}{\textbf{Smell}}
& \multicolumn{8}{!{\vrule width 0.35pt}c!{\vrule width 0.35pt}}{\textbf{GPT-5.4-nano}}
& \multicolumn{8}{c!{\vrule width 0.35pt}}{\textbf{Gemini 3.5 Flash}}
& \multicolumn{8}{c!{\vrule width 0.35pt}}{\textbf{Claude Haiku 4.5}}
& \multicolumn{8}{c}{\textbf{DeepSeek V4 Flash}} \\

\cline{3-34}

&
& \multicolumn{4}{!{\vrule width 0.35pt}c!{\vrule width 0.25pt}}{\textbf{D1}}
& \multicolumn{4}{c!{\vrule width 0.35pt}}{\textbf{D2}}
& \multicolumn{4}{c!{\vrule width 0.25pt}}{\textbf{D1}}
& \multicolumn{4}{c!{\vrule width 0.35pt}}{\textbf{D2}}
& \multicolumn{4}{c!{\vrule width 0.25pt}}{\textbf{D1}}
& \multicolumn{4}{c!{\vrule width 0.35pt}}{\textbf{D2}}
& \multicolumn{4}{c!{\vrule width 0.25pt}}{\textbf{D1}}
& \multicolumn{4}{c}{\textbf{D2}} \\

\cline{3-34}

&
& \textbf{F1} & \textbf{Pr} & \textbf{Re} & \textbf{Acc}
& \textbf{F1} & \textbf{Pr} & \textbf{Re} & \textbf{Acc}
& \textbf{F1} & \textbf{Pr} & \textbf{Re} & \textbf{Acc}
& \textbf{F1} & \textbf{Pr} & \textbf{Re} & \textbf{Acc}
& \textbf{F1} & \textbf{Pr} & \textbf{Re} & \textbf{Acc}
& \textbf{F1} & \textbf{Pr} & \textbf{Re} & \textbf{Acc}
& \textbf{F1} & \textbf{Pr} & \textbf{Re} & \textbf{Acc}
& \textbf{F1} & \textbf{Pr} & \textbf{Re} & \textbf{Acc} \\
\midrule

& \texttt{MissingPermExternalActions}  & \underline{97} & 100 & 93 & 97 & \textbf{100} & 100 & 100 & 100 & \textbf{100} & 100 & 100 & 100 & \textbf{100} & 100 & 100 & 100 & \textbf{100} & 100 & 100 & 100 & \textbf{100} & 100 & 100 & 100 & \textbf{100} & 100 & 100 & 100 & \textbf{100} & 100 & 100 & 100 \\

\rowcolor{gray!20}
\cellcolor{white}
& \texttt{UnpinnedActionVersion} & \textbf{80} & 100 & 67 & 83 & \textbf{80} & 100 & 67 & 83 & \textbf{80} & 100 & 67 & 83 & \textbf{80} & 100 & 67 & 83 & \textbf{80} & 100 & 67 & 83 & \textbf{80} & 100 & 67 & 83 & \textbf{80} & 100 & 67 & 83 & \textbf{80} & 100 & 67 & 83 \\

& \texttt{MissingTokenPermissions} & \underline{97} & 100 & 93 & 97 & \textbf{100} & 100 & 100 & 100 & \textbf{100} & 100 & 100 & 100 & \textbf{100} & 100 & 100 & 100 & \textbf{100} & 100 & 100 & 100 & \textbf{100} & 100 & 100 & 100 & \textbf{100} & 100 & 100 & 100 & \textbf{100} & 100 & 100 & 100 \\

\rowcolor{gray!20}
\cellcolor{white}
& \texttt{MissingJobTimeout} & \textbf{82} & 74 & 93 & 80 & \underline{75} & 60 & 100 & 67 & \textbf{100} & 100 & 100 & 100 & \underline{79} & 65 & 100 & 73 & \textbf{75} & 60 & 100 & 67 & \textbf{75} & 60 & 100 & 67 & \textbf{77} & 62 & 100 & 70 & \underline{75} & 60 & 100 & 67 \\

& \texttt{UnsafeArtifactUploadFork} & \underline{85} & 100 & 73 & 87 & \textbf{100} & 100 & 100 & 100 & \textbf{100} & 100 & 100 & 100 & \textbf{100} & 100 & 100 & 100 & \textbf{100} & 100 & 100 & 100 & \textbf{100} & 100 & 100 & 100 & \textbf{100} & 100 & 100 & 100 & \textbf{100} & 100 & 100 & 100 \\

& \texttt{MissingUploadCondition} & \underline{70} & 100 & 53 & 77 & \textbf{93} & 100 & 87 & 93 & \underline{89} & 100 & 80 & 90 & \textbf{93} & 100 & 87 & 93 & \textbf{93} & 100 & 87 & 93 & \textbf{93} & 100 & 87 & 93 & \underline{89} & 100 & 80 & 90 & \textbf{93} & 100 & 87 & 93 \\

& \texttt{MutableRunnerVersion} & \underline{89} & 100 & 80 & 90 & \textbf{93} & 100 & 87 & 93 & \textbf{93} & 100 & 87 & 93 & \textbf{93} & 100 & 87 & 93 & \textbf{89} & 100 & 80 & 90 & \textbf{89} & 100 & 80 & 90 & \textbf{93} & 100 & 87 & 93 & \underline{89} & 100 & 80 & 90 \\

& \texttt{SuboptimalCacheUsage} & \underline{70} & 100 & 53 & 77 & \textbf{100} & 100 & 100 & 100 & \textbf{100} & 100 & 100 & 100 & \textbf{100} & 100 & 100 & 100 & \textbf{100} & 100 & 100 & 100 & \textbf{100} & 100 & 100 & 100 & \textbf{100} & 100 & 100 & 100 & \textbf{100} & 100 & 100 & 100 \\

\rowcolor{gray!20}
\cellcolor{white}
& \texttt{MultiCommandStep} & \underline{70} & 100 & 53 & 77 & \textbf{80} & 100 & 67 & 83 & \underline{57} & 100 & 40 & 70 & \textbf{64} & 100 & 47 & 73 & \textbf{70} & 100 & 53 & 77 & \underline{64} & 100 & 47 & 73 & \textbf{70} & 100 & 53 & 77 & \underline{64} & 100 & 47 & 73 \\

\multirow{-10}{*}{\groupcell{\textbf{Localized}}}
& \texttt{UnpinnedDependencies} & \textbf{100} & 100 & 100 & 100 & \textbf{100} & 100 & 100 & 100 & \underline{80} & 100 & 67 & 83 & \textbf{89} & 100 & 80 & 90 & \textbf{97} & 100 & 93 & 97 & \textbf{97} & 100 & 93 & 97 & \textbf{97} & 100 & 93 & 97 & \textbf{97} & 100 & 93 & 97 \\

\midrule

\rowcolor{gray!20}
\cellcolor{white}
& \texttt{MissingForkGuardPR} & \textbf{42} & 100 & 27 & 63 & \textbf{42} & 100 & 27 & 63 & \textbf{42} & 100 & 27 & 63 & \underline{24} & 100 & 13 & 57 & \textbf{64} & 100 & 47 & 73 & \underline{50} & 100 & 33 & 67 & \textbf{42} & 100 & 27 & 63 & \textbf{42} & 100 & 27 & 63 \\

& \texttt{MissingConcurrencyPR} & \textbf{100} & 100 & 100 & 100 & \underline{97} & 100 & 93 & 97 & \textbf{100} & 100 & 100 & 100 & \textbf{100} & 100 & 100 & 100 & \textbf{100} & 100 & 100 & 100 & \textbf{100} & 100 & 100 & 100 & \textbf{100} & 100 & 100 & 100 & \textbf{100} & 100 & 100 & 100 \\

& \texttt{MissingConcurrencyBranch} & \textbf{87} & 87 & 87 & 87 & \underline{80} & 80 & 80 & 80 & \textbf{94} & 88 & 100 & 93 & \underline{91} & 83 & 100 & 90 & \textbf{94} & 88 & 100 & 93 & \underline{91} & 83 & 100 & 90 & \textbf{91} & 83 & 100 & 90 & \textbf{91} & 83 & 100 & 90 \\

& \texttt{MissingPathFilters} & \underline{97} & 100 & 93 & 97 & \textbf{100} & 100 & 100 & 100 & \underline{97} & 100 & 93 & 97 & \textbf{100} & 100 & 100 & 100 & \textbf{97} & 100 & 93 & 97 & \underline{93} & 100 & 87 & 93 & \textbf{97} & 100 & 93 & 97 & \textbf{97} & 100 & 93 & 97 \\

& \texttt{MissingForkGuardSchedule} & \underline{42} & 100 & 27 & 63 & \textbf{80} & 100 & 67 & 83 & \textbf{100} & 100 & 100 & 100 & \textbf{100} & 100 & 100 & 100 & \textbf{100} & 100 & 100 & 100 & \textbf{100} & 100 & 100 & 100 & \textbf{100} & 100 & 100 & 100 & \textbf{100} & 100 & 100 & 100 \\

& \texttt{UnsafeDeployOnFork} & \underline{75} & 100 & 60 & 80 & \textbf{89} & 100 & 80 & 90 & \textbf{89} & 100 & 80 & 90 & \textbf{89} & 100 & 80 & 90 & \textbf{89} & 100 & 80 & 90 & \textbf{89} & 100 & 80 & 90 & \textbf{85} & 100 & 73 & 87 & \textbf{85} & 100 & 73 & 87 \\

& \texttt{MissingConcurrencyGlobal} & \textbf{86} & 92 & 80 & 87 & \underline{76} & 79 & 73 & 77 & \textbf{79} & 85 & 73 & 80 & \underline{58} & 78 & 47 & 67 & \underline{88} & 79 & 100 & 87 & \textbf{90} & 88 & 93 & 90 & \textbf{90} & 88 & 93 & 90 & \underline{88} & 79 & 100 & 87 \\

\rowcolor{gray!20}
\cellcolor{white}
& \texttt{SingleOSExecution} & \underline{64} & 100 & 47 & 73 & \textbf{67} & 89 & 53 & 73 & \underline{24} & 100 & 13 & 57 & \textbf{61} & 88 & 47 & 70 & \underline{48} & 83 & 33 & 63 & \textbf{67} & 89 & 53 & 73 & \textbf{67} & 89 & 53 & 73 & \underline{61} & 88 & 47 & 70 \\

\multirow{-9}{*}{
    \rotatebox[origin=c]{90}{
        \shortstack{
            \textsc{\scriptsize \textbf{Context-}}\\
            \textsc{\scriptsize \textbf{dependent}}
        }
    }
}
& \texttt{SingleLangVersionCI} & \underline{75} & 100 & 60 & 80 & \textbf{86} & 92 & 80 & 87 & \underline{70} & 100 & 53 & 77 & \textbf{80} & 100 & 67 & 83 & \textbf{86} & 92 & 80 & 87 & \underline{85} & 100 & 73 & 87 & \underline{75} & 100 & 60 & 80 & \textbf{85} & 100 & 73 & 87 \\

\midrule
\multicolumn{2}{l!{\vrule width 0.35pt}}{\textbf{Avg.}}
& \textbf{79} & \textbf{98} & \textbf{71} & \textbf{84}
& \textbf{86} & \textbf{95} & \textbf{82} & \textbf{88}
& \textbf{84} & \textbf{99} & \textbf{78} & \textbf{88}
& \textbf{84} & \textbf{95} & \textbf{80} & \textbf{88}
& \textbf{88} & \textbf{95} & \textbf{85} & \textbf{89}
& \textbf{87} & \textbf{96} & \textbf{84} & \textbf{89}
& \textbf{87} & \textbf{96} & \textbf{83} & \textbf{89}
& \textbf{87} & \textbf{95} & \textbf{84} & \textbf{89} \\
\bottomrule
\end{tabular}
\end{adjustbox}
\end{table*}

In contrast, several smells remain more challenging. Context-dependent smells such as \textit{MissingForkGuardPR}, \textit{SingleOSExecution}, and \textit{MissingConcurrencyGlobal} show lower or more variable F1 scores, reflecting the difficulty of recovering true positives when evidence depends on fork context, concurrency semantics, or OS coverage. The localized but intent-sensitive \textit{MultiCommandStep} is also difficult, ranging from 57\% to 80\% F1. A different pattern appears for \textit{UnpinnedActionVersion}: all models and configurations are conservative, with 100\% precision but only 67\% recall. These results suggest that detection becomes harder when models must infer workflow intent, distinguish separable command phases, or consistently apply strict pinning criteria to action references.

We find that task decomposition helps mostly when the model under-detects in D1. GPT-5.4-nano benefits the most from D2, improving from 79\% to 86\% in smell-level macro-F1, mainly through recall gains from 71\% to 82\%. These gains are visible for \textit{SuboptimalCacheUsage}, \textit{MissingUploadCondition}, \textit{MissingForkGuardSchedule}, \textit{UnsafeArtifactUploadFork}, and \textit{SingleLangVersionCI}. However, this pattern does not generalize to all models: Gemini 3.5 Flash remains stable on average, while Claude Haiku 4.5 and DeepSeek V4 Flash are already strong under D1 and change little under D2.
Although the open-weight DeepSeek V4 Flash remains competitive with closed models, it benefits little from decomposition, whereas GPT-5.4-nano benefits substantially. Thus, decomposition gains are not explained by model availability alone, but by how each model handles task decomposition and smell-specific reasoning demands.

\begin{finding}
Detection gains are selective: D2 mainly helps GPT-5.4-nano and a subset of target smells, while models with stronger D1 performance change little under decomposition.
\end{finding}

\subsection{Refactoring Outcomes (RQ$_2$)}
\label{sec:rq2}

To answer \textbf{RQ$_2$}, we evaluate whether the studied \textsc{GHA-SmellFix} orchestration designs generate structurally valid and scope-bounded candidate refactorings for workflow smells. 

\textbf{End-to-end construct-aware outcomes by target smell.}
Following the refactoring assessment procedure described in Step~6, we classify positive-pair outputs as Clean Removal (CR), Removal with Unexpected Changes (UX), Removal with Potential Behavior-Impacting Change (BR), or Failed/No Patch (FN). For negative pairs, we distinguish Correct Abstention (CA) from Unnecessary Edit (UE). Table~\ref{tab:rq2_refactoring_outcomes_by_smell} reports these end-to-end outcomes by smell, together with the Smell Removal Rate (SRR = CR + UX + BR). Because FN conflates missed detections and failed patches after detection, we separate them in the conditional analysis below. Arrows indicate the desired direction. 

\begin{table*}[ht!]
\centering
\caption{Construct-aware refactoring outcomes by target smell. Values aggregate models and designs; percentages include counts in parentheses. Gray rows highlight the lowest-CR smells.}

\label{tab:rq2_refactoring_outcomes_by_smell}
\resizebox{\textwidth}{!}{%
\begin{tabular}{llrrrrrl|rrl}
\toprule
\multirow{2}{*}{\textbf{Type}} &
  \multirow{2}{*}{\textbf{Smell}} &
  \multicolumn{6}{c|}{\textbf{Ground-truth-positive outputs}} &
  \multicolumn{3}{c}{\textbf{Ground-truth-negative outputs}} \\
\cmidrule(lr){3-8}
\cmidrule(lr){9-11}
 &
 &
  \textbf{\begin{tabular}[c]{r}SRR $\uparrow$\end{tabular}} &
  \textbf{\begin{tabular}[c]{r}CR $\uparrow$\end{tabular}} &
  \textbf{\begin{tabular}[c]{r}UX $\downarrow$\end{tabular}} &
  \textbf{\begin{tabular}[c]{r}BR $\downarrow$\end{tabular}} &
  \textbf{\begin{tabular}[c]{r}FN $\downarrow$\end{tabular}} &
  \textbf{\begin{tabular}[c]{l}Assessable/Expected\\ Pos. Outputs\end{tabular}} &
  \textbf{\begin{tabular}[c]{r}CA $\uparrow$\end{tabular}} &
  \textbf{\begin{tabular}[c]{r}UE $\downarrow$\end{tabular}} &
  \textbf{\begin{tabular}[c]{l}Assessable/Expected\\ Negative Outputs\end{tabular}} \\
\midrule

\multirow{10}{*}{
\rotatebox[origin=c]{90}{
        \shortstack{
            \textsc{ \textbf{Localized}}}
        }
    }
     &
  \texttt{MissingPermExternalActions} &
  ~99\%~{\scriptsize(117)} &
  ~98\%~{\scriptsize(116)} &
  ~1\%~{\scriptsize(1)} &
  ~0\%~{\scriptsize(0)} &
  ~1\%~{\scriptsize(1)} &
  ~118/120 &
  ~100\%~{\scriptsize(120)} &
  ~0\%~{\scriptsize(0)} &
  ~120/120 \\

  &
  \cellcolor{gray!10}\texttt{UnpinnedActionVersion} &
  \cellcolor{gray!10}~66\%~{\scriptsize(78)} &
  \cellcolor{gray!10}~40\%~{\scriptsize(47)} &
  \cellcolor{gray!10}~0\%~{\scriptsize(0)} &
  \cellcolor{gray!10}~26\%~{\scriptsize(31)} &
  \cellcolor{gray!10}~34\%~{\scriptsize(40)} &
  \cellcolor{gray!10}~118/120 &
  \cellcolor{gray!10}~100\%~{\scriptsize(120)} &
  \cellcolor{gray!10}~0\%~{\scriptsize(0)} &
  \cellcolor{gray!10}~120/120 \\

  &
  \texttt{MissingTokenPermissions} &
  ~99\%~{\scriptsize(119)} &
  ~99\%~{\scriptsize(119)} &
  ~0\%~{\scriptsize(0)} &
  ~0\%~{\scriptsize(0)} &
  ~1\%~{\scriptsize(1)} &
  ~120/120 &
  ~100\%~{\scriptsize(120)} &
  ~0\%~{\scriptsize(0)} &
  ~120/120 \\

  &
  \texttt{MissingJobTimeout} &
  ~93\%~{\scriptsize(105)} &
  ~91\%~{\scriptsize(103)} &
  ~0\%~{\scriptsize(0)} &
  ~2\%~{\scriptsize(2)} &
  ~7\%~{\scriptsize(8)} &
  ~113/120 &
  ~56\%~{\scriptsize(65)} &
  ~44\%~{\scriptsize(52)} &
  ~117/120 \\

  &
  \texttt{UnsafeArtifactUploadFork} &
  ~95\%~{\scriptsize(110)} &
  ~95\%~{\scriptsize(110)} &
  ~0\%~{\scriptsize(0)} &
  ~0\%~{\scriptsize(0)} &
  ~5\%~{\scriptsize(6)} &
  ~116/120 &
  ~100\%~{\scriptsize(120)} &
  ~0\%~{\scriptsize(0)} &
  ~120/120 \\

  &
  \texttt{MissingUploadCondition} &
  ~77\%~{\scriptsize(75)} &
  ~77\%~{\scriptsize(75)} &
  ~0\%~{\scriptsize(0)} &
  ~0\%~{\scriptsize(0)} &
  ~23\%~{\scriptsize(23)} &
  ~98/120 &
  ~100\%~{\scriptsize(120)} &
  ~0\%~{\scriptsize(0)} &
  ~120/120 \\

  &
  \texttt{MutableRunnerVersion} &
  ~82\%~{\scriptsize(96)} &
  ~81\%~{\scriptsize(95)} &
  ~0\%~{\scriptsize(0)} &
  ~1\%~{\scriptsize(1)} &
  ~18\%~{\scriptsize(21)} &
  ~117/120 &
  ~100\%~{\scriptsize(120)} &
  ~0\%~{\scriptsize(0)} &
  ~120/120 \\

  &
  \texttt{SuboptimalCacheUsage} &
  ~91\%~{\scriptsize(105)} &
  ~86\%~{\scriptsize(99)} &
  ~1\%~{\scriptsize(1)} &
  ~4\%~{\scriptsize(5)} &
  ~9\%~{\scriptsize(10)} &
  ~115/120 &
  ~100\%~{\scriptsize(120)} &
  ~0\%~{\scriptsize(0)} &
  ~120/120 \\

  &
  \cellcolor{gray!10}\texttt{MultiCommandStep} &
  \cellcolor{gray!10}~47\%~{\scriptsize(54)} &
  \cellcolor{gray!10}~46\%~{\scriptsize(53)} &
  \cellcolor{gray!10}~0\%~{\scriptsize(0)} &
  \cellcolor{gray!10}~1\%~{\scriptsize(1)} &
  \cellcolor{gray!10}~53\%~{\scriptsize(62)} &
  \cellcolor{gray!10}~116/120 &
  \cellcolor{gray!10}~100\%~{\scriptsize(120)} &
  \cellcolor{gray!10}~0\%~{\scriptsize(0)} &
  \cellcolor{gray!10}~120/120 \\

  &
  \texttt{UnpinnedDependencies} &
  ~89\%~{\scriptsize(102)} &
  ~70\%~{\scriptsize(81)} &
  ~1\%~{\scriptsize(1)} &
  ~17\%~{\scriptsize(20)} &
  ~11\%~{\scriptsize(13)} &
  ~115/120 &
  ~100\%~{\scriptsize(120)} &
  ~0\%~{\scriptsize(0)} &
  ~120/120 \\

\midrule

\multirow{9}{*}{\rotatebox[origin=c]{90}{
        \shortstack{
            \textsc{ \textbf{Context-}}\\
            \textsc{ \textbf{dependent}}
        }
    }} &
  \cellcolor{gray!10}\texttt{MissingForkGuardPR} &
  \cellcolor{gray!10}~26\%~{\scriptsize(30)} &
  \cellcolor{gray!10}~25\%~{\scriptsize(29)} &
  \cellcolor{gray!10}~1\%~{\scriptsize(1)} &
  \cellcolor{gray!10}~0\%~{\scriptsize(0)} &
  \cellcolor{gray!10}~74\%~{\scriptsize(87)} &
  \cellcolor{gray!10}~117/120 &
  \cellcolor{gray!10}~100\%~{\scriptsize(120)} &
  \cellcolor{gray!10}~0\%~{\scriptsize(0)} &
  \cellcolor{gray!10}~120/120 \\

  &
  \texttt{MissingConcurrencyPR} &
  ~99\%~{\scriptsize(118)} &
  ~97\%~{\scriptsize(115)} &
  ~3\%~{\scriptsize(3)} &
  ~0\%~{\scriptsize(0)} &
  ~1\%~{\scriptsize(1)} &
  ~119/120 &
  ~100\%~{\scriptsize(120)} &
  ~0\%~{\scriptsize(0)} &
  ~120/120 \\

  &
  \texttt{MissingConcurrencyBranch} &
  ~96\%~{\scriptsize(115)} &
  ~96\%~{\scriptsize(115)} &
  ~0\%~{\scriptsize(0)} &
  ~0\%~{\scriptsize(0)} &
  ~4\%~{\scriptsize(5)} &
  ~120/120 &
  ~88\%~{\scriptsize(100)} &
  ~12\%~{\scriptsize(14)} &
  ~114/120 \\

  &
  \texttt{MissingPathFilters} &
  ~91\%~{\scriptsize(107)} &
  ~81\%~{\scriptsize(95)} &
  ~9\%~{\scriptsize(10)} &
  ~2\%~{\scriptsize(2)} &
  ~9\%~{\scriptsize(10)} &
  ~117/120 &
  ~100\%~{\scriptsize(120)} &
  ~0\%~{\scriptsize(0)} &
  ~120/120 \\

  &
  \texttt{MissingForkGuardSchedule} &
  ~84\%~{\scriptsize(91)} &
  ~81\%~{\scriptsize(88)} &
  ~3\%~{\scriptsize(3)} &
  ~0\%~{\scriptsize(0)} &
  ~16\%~{\scriptsize(17)} &
  ~108/120 &
  ~100\%~{\scriptsize(120)} &
  ~0\%~{\scriptsize(0)} &
  ~120/120 \\

  &
  \texttt{UnsafeDeployOnFork} &
  ~73\%~{\scriptsize(86)} &
  ~70\%~{\scriptsize(83)} &
  ~1\%~{\scriptsize(1)} &
  ~2\%~{\scriptsize(2)} &
  ~27\%~{\scriptsize(32)} &
  ~118/120 &
  ~100\%~{\scriptsize(120)} &
  ~0\%~{\scriptsize(0)} &
  ~120/120 \\

  &
  \texttt{MissingConcurrencyGlobal} &
  ~75\%~{\scriptsize(85)} &
  ~74\%~{\scriptsize(84)} &
  ~1\%~{\scriptsize(1)} &
  ~0\%~{\scriptsize(0)} &
  ~25\%~{\scriptsize(28)} &
  ~113/120 &
  ~83\%~{\scriptsize(100)} &
  ~17\%~{\scriptsize(20)} &
  ~120/120 \\

  &
  \cellcolor{gray!10}\texttt{SingleOSExecution} &
  \cellcolor{gray!10}~37\%~{\scriptsize(41)} &
  \cellcolor{gray!10}~30\%~{\scriptsize(33)} &
  \cellcolor{gray!10}~0\%~{\scriptsize(0)} &
  \cellcolor{gray!10}~7\%~{\scriptsize(8)} &
  \cellcolor{gray!10}~63\%~{\scriptsize(69)} &
  \cellcolor{gray!10}~110/120 &
  \cellcolor{gray!10}~95\%~{\scriptsize(114)} &
  \cellcolor{gray!10}~5\%~{\scriptsize(6)} &
  \cellcolor{gray!10}~120/120 \\

  &
  \cellcolor{gray!10}\texttt{SingleLangVersionCI} &
  \cellcolor{gray!10}~65\%~{\scriptsize(73)} &
  \cellcolor{gray!10}~65\%~{\scriptsize(73)} &
  \cellcolor{gray!10}~0\%~{\scriptsize(0)} &
  \cellcolor{gray!10}~0\%~{\scriptsize(0)} &
  \cellcolor{gray!10}~35\%~{\scriptsize(40)} &
  \cellcolor{gray!10}~113/120 &
  \cellcolor{gray!10}~99\%~{\scriptsize(119)} &
  \cellcolor{gray!10}~1\%~{\scriptsize(1)} &
  \cellcolor{gray!10}~120/120 \\

\midrule

\multicolumn{2}{l}{\textbf{Overall}} &
  \textbf{~78\%~{\scriptsize(1,707)}} &
  \textbf{~74\%~{\scriptsize(1,613)}} &
  \textbf{~1\%~{\scriptsize(22)}} &
  \textbf{~3\%~{\scriptsize(72)}} &
  \textbf{~22\%~{\scriptsize(474)}} &
  \textbf{~2,181/2,280} &
  \textbf{~96\%~{\scriptsize(2,178)}} &
  \textbf{~4\%~{\scriptsize(93)}} &
  \textbf{~2,271/2,280} \\

\bottomrule
\end{tabular}%
}

\end{table*}

Table~\ref{tab:rq2_refactoring_outcomes_by_smell} shows that clean removal is often high 
for smells whose fixes target explicit workflow constructs, such as \textit{MissingPermExternalActions}, \textit{MissingConcurrencyPR}, and \textit{UnsafeArtifactUploadFork}. However, two limitations remain. First, high SRR does not always imply clean refactoring: \textit{UnpinnedActionVersion} and \textit{UnpinnedDependencies} show SRR--CR gaps mainly due to potential behavior-impacting changes, while \textit{MissingPathFilters} shows a smaller gap due mostly to unexpected changes. Second, difficult cases are not limited to one group: context-dependent smells such as \textit{MissingForkGuardPR} and \textit{SingleOSExecution}, as well as the localized but intent-sensitive \textit{MultiCommandStep}, remain difficult, with low CR and high FN. 

Figure~\ref{fig:rq2_examples} illustrates these construct-aware distinctions, showing that patches may be clean, include collateral edits, depend on external facts, or modify workflows where the target smell is absent.

\begin{figure}[ht!]
\scriptsize
\centering
\begin{adjustbox}{width=\columnwidth}
% ---- linha 1 ----
\begin{minipage}[t]{0.42\linewidth}\centering
{\tiny\textbf{(a) Clean removal}}\par\vspace{-1pt}
\begin{lstlisting}[
numbers=left,           % Posição dos números: left, right ou none
numberstyle=\tiny,      % Tamanho da fonte dos números
stepnumber=1,           % Exibe número a cada X linhas
numbersep=5pt,          % Distância entre o número e o código
columns=fullflexible,
basicstyle=\ttfamily\tiny,
breaklines=true,
language=diff,
xleftmargin=0.2em]

--- original
+++ patch
@@ -6,6 +6,9 @@
     branches:
       - master
       - release-*
+concurrency: (*\anno{1}*)(*\fixmark*)
+  group: ${{ github.workflow }}-${{ github.ref }}
+  cancel-in-progress: true
 permissions:
   contents: read
 jobs:
\end{lstlisting}
\end{minipage}
\hfill
\hfill

\begin{minipage}[t]{0.47\linewidth}\centering
{\tiny\textbf{(b) Unexpected changes}}\par\vspace{-1pt}
\begin{lstlisting}[
numbers=left,           % Posição dos números: left, right ou none
numberstyle=\tiny,      % Tamanho da fonte dos números
stepnumber=1,           % Exibe número a cada X linhas
numbersep=5pt,          % Distância entre o número e o código
columns=fullflexible,
basicstyle=\ttfamily\tiny,
breaklines=true,
language=diff,
xleftmargin=0.2em]

--- original
+++ patch
@@ -10,14 +10,17 @@
 permissions:
   contents: read
-  id-token: "write"  # AWS GitHub OIDC required: write (*\anno{2}*)
+  id-token: "write"
   packages: read
+concurrency: (*\anno{1}*)(*\fixmark*)
+  group: ${{ github.workflow }}-${{ github.event.pull_request.number || github.ref }}
+  cancel-in-progress: true
 jobs:
   flowzone:
     name: Flowzone
     uses: product-os/flowzone/.github/workflows/flowzone.yml@master
-    # prevent duplicate workflow executions for pull_request and pull_request_target (*\anno{3}*)
     if: |
\end{lstlisting}
\end{minipage}
\hfill
\hfill
% ---- linha 2 ----
\begin{minipage}[t]{0.50\linewidth}\centering
{\tiny\textbf{(c) Potential behavior-impacting change}}\par\vspace{-1pt}
\begin{lstlisting}[
numbers=left,           % Posição dos números: left, right ou none
numberstyle=\tiny,      % Tamanho da fonte dos números
stepnumber=1,           % Exibe número a cada X linhas
numbersep=5pt,          % Distância entre o número e o código
columns=fullflexible,
basicstyle=\ttfamily\tiny,
breaklines=true,
language=diff,
xleftmargin=0.2em]

--- original
+++ patch
@@ -21,7 +21,7 @@
     timeout-minutes: 5
     steps:
       - name: Run microbenchmark
-        uses: elastic/oblt-actions/buildkite/run@v1 (*\anno{1}*)(*\smellmark*)
+        uses: elastic/oblt-actions/buildkite/run@a1b2c3...a1b (*\anno{2}*)(*\fixmark*)
        with:
          pipeline: "apm-agent-microbenchmark"
          token: ${{ secrets.BUILDKITE_TOKEN }}
\end{lstlisting}
\end{minipage}\hfill
% https://github.com/meta-pytorch/torchrec/blob/3bdf9f384ce6a0ead822cdd6780d8807c760260c/.github/workflows/unittest_ci.yml
\begin{minipage}[t]{0.38\linewidth}\centering
{\tiny\textbf{(d) Unnecessary (neg.)}}\par\vspace{-1pt}
\begin{lstlisting}[
numbers=left,           % Posição dos números: left, right ou none
numberstyle=\tiny,      % Tamanho da fonte dos números
stepnumber=1,           % Exibe número a cada X linhas
numbersep=5pt,          % Distância entre o número e o código
columns=fullflexible,
basicstyle=\ttfamily\tiny,
breaklines=true,
language=diff,
xleftmargin=0.2em]

--- original
+++ patch
@@ -1,6 +1,3 @@
-# This workflow will install Python dependencies... (*\anno{1}*)
@@ -41,6 +38,7 @@
 jobs:
   unittest_ci_gpu:
+    timeout-minutes: 60 (*\anno{2}*)
     strategy:
@@ -91,7 +89,7 @@
             python:
-              version: "3.11"
+              version: "3.10" (*\anno{3}*)
@@ -125,12 +123,15 @@
-    uses: pytorch/test-infra/.github/workflows/linux_job_v2.yml@main (*\anno{4}*)
...
+    steps: (*\anno{5}*)
+      - name: Checkout
+        uses: actions/checkout@v4
         ...
\end{lstlisting}
\end{minipage}
\end{adjustbox}
\caption{Representative refactoring outcomes: clean removal (CR), removal with unexpected changes (UX), removal with potential behavior-impacting change (BR), and unnecessary edit on a negative pair (UE). Markers denote the target smell construct (\smellmark{}), the intended fix (\fixmark{}), and other edits (\anno{n}).}
\label{fig:rq2_examples}
\end{figure}

Figure~\ref{fig:rq2_examples}(a) shows a clean-removal case for \textit{MissingConcurrencyBranch}. In the inspected outputs for this workflow, Gemini 3.5 Flash produces the same construct-level modification under D1 and D2, and DeepSeek V4 Flash shows the same behavior. Although placement or formatting may vary across outputs, the modified construct is the same: a single \texttt{concurrency} block added at marker~\anno{1} in \faLink~\href{https://github.com/kumahq/kuma/blob/5067a4e3192f44baa22f9afa652a1f1ac85b59b5/.github/workflows/pr-merged.yaml}{\texttt{pr-merged.yaml}}. The added \texttt{group} scopes concurrent runs by workflow and reference, while \texttt{cancel-in-progress: true} cancels older in-flight runs when a newer run in the same group is triggered. Because the patch is purely additive and touches no existing job or step, it is classified as a clean removal. This helps explain the high CR and 0\% UX for \textit{MissingConcurrencyBranch} in Table~\ref{tab:rq2_refactoring_outcomes_by_smell}; the next examples show why smell removal alone is insufficient to assess patch quality.

Figure~\ref{fig:rq2_examples}(b) shows a \textit{removal-with-unexpected-changes case} for \textit{MissingConcurrencyPR} in the \faLink~\href{https://github.com/balena-io/open-balena/blob/8d8c5e2c8d2ece3641e2edb52720b6cba1a4cf89/.github/workflows/flowzone.yml}{\texttt{flowzone.yml}}. All eight outputs for this workflow, from the four models under D1 and D2, converge on the same construct-level fix: a top-level \texttt{concurrency} block with \texttt{cancel-in-progress: true}, added at marker~\anno{1}; only the \texttt{group} key varies. In D1, however, DeepSeek V4 Flash also deletes two comments outside the fix: the note on why \texttt{id-token: write} is required, at marker~\anno{2}, and the note on the purpose of the job-level \texttt{if} guard, at marker~\anno{3}. These deletions do not change how the workflow runs, but they remove documentation of the existing configuration. Thus, the intended fix is present, but the patch is not clean. The other seven outputs apply the same fix without collateral edits, so the collateral edit comes from one output, not from the fix itself. This helps explain why \textit{MissingConcurrencyPR} still has 3\% UX despite its 97\% CR in Table~\ref{tab:rq2_refactoring_outcomes_by_smell}.

Figure~\ref{fig:rq2_examples}(c) shows a \textit{potential behavior-impacting change}. In D1, Claude Haiku 4.5 refactors \textit{UnpinnedActionVersion} by replacing the mutable \texttt{@v1} tag in the action-reference construct at marker~\anno{1} with a 40-character commit-like reference at marker~\anno{2} in \faLink~\href{https://github.com/elastic/apm-agent-nodejs/blob/6e047fefe0f95973fccdf7698506c5a674a81c24/.github/workflows/microbenchmark.yml}{\texttt{microbenchmark.yml}}. Structurally, this is the expected fix: the action reference is no longer pinned to a mutable version tag. However, correctness depends on a fact outside the workflow file, namely whether the generated reference corresponds to a real commit in the action repository. In this example, the \texttt{a1b2c3...} pattern suggests a placeholder-like value, so the patch is well-formed but requires repository-level validation before use. Thus, the smell is removed syntactically, but the patch introduces a potential behavior-impacting change because it may break the step or pin an unintended revision. 
This explains why \textit{UnpinnedActionVersion} has low CR and non-trivial BR in Table~\ref{tab:rq2_refactoring_outcomes_by_smell} (40\% CR, 26\% BR).

Figure~\ref{fig:rq2_examples}(d) shows an \textit{unnecessary-edit} case on a negative pair, where \textit{MissingJobTimeout} is absent and abstention is expected. In D1, GPT-5.4-nano nevertheless edits in the \faLink~\href{https://github.com/meta-pytorch/torchrec/blob/3bdf9f384ce6a0ead822cdd6780d8807c760260c/.github/workflows/unittest_ci.yml}{\texttt{unittest\_ci.yml}}: it adds \texttt{timeout-minutes} at marker~\anno{2}, but also changes unrelated constructs, removing a header comment at marker~\anno{1}, changing the Python matrix from \texttt{3.11} to \texttt{3.10} at marker~\anno{3}, and replacing a reusable-workflow call at marker~\anno{4} with inline \texttt{steps} at marker~\anno{5}. This is not a failed refactoring, but an unnecessary rewrite of a workflow that should have remained unchanged. 
Apart from comment deletion, edits are behavior-relevant rather than cosmetic, since they may alter job duration, runtime environment, and workflow reuse.
This helps explain why \textit{MissingJobTimeout} is the smell most prone to unnecessary edits, with a 44\% UE rate on negative pairs.

\begin{finding}
Smell removal overestimates refactoring quality. Some smell-removing patches still introduce unexpected edits or potential behavior-impacting changes, so CI/CD refactoring evaluation should distinguish smell removal from construct-aware patch quality.

\end{finding}

\textbf{End-to-end failures are mostly missed detections.}
Table~\ref{tab:rq2_end_to_end_conditional} reports end-to-end (E2E) and conditional (Cond.) outcomes; Det. miss and Patch fail denote detection misses and post-detection patch failures. End-to-end, the pipelines remove the target smell in 78.3\% of assessable outputs and produce clean removals in 74\%. Among the 474 \textit{Failed/No Patch} outcomes, 439 (92.6\%) are detection misses and only 35 are post-detection patch failures.

\begingroup
\setlength{\columnsep}{6pt}   % espaço horizontal entre texto e tabela
\setlength{\intextsep}{4pt}  
\begin{wraptable}{r}{7.5cm}
\centering
\caption{End-to-end and conditional refactoring outcomes on positive outputs.}
\label{tab:rq2_end_to_end_conditional}
\scriptsize
\setlength{\tabcolsep}{2.2pt}
\renewcommand{\arraystretch}{1.02}
\begin{tabular}{llrrrrr}
\toprule
\textbf{Perspective} & \textbf{Design} & \textbf{Out.} &
\textbf{\begin{tabular}[c]{@{}r@{}}Det.\\miss\end{tabular}} &
\textbf{\begin{tabular}[c]{@{}r@{}}Patch\\fail\end{tabular}} &
\textbf{SRR $\uparrow$} & \textbf{CR $\uparrow$} \\
\midrule
\multirow{3}{*}{E2E}
& D1 & 1,092 & 238 & 21 & 76.3\% {\scriptsize(833)} & 71.8\% {\scriptsize(784)} \\
& D2 & 1,089 & 201 & 14 & 80.3\% {\scriptsize(874)} & 76.1\% {\scriptsize(829)} \\
\cmidrule(lr){2-7}
& All & 2,181 & 439 & 35 & 78.3\% {\scriptsize(1,707)} & 74.0\% {\scriptsize(1,613)} \\
\midrule
\multirow{3}{*}{Cond.}
& D1 & 854 & -- & 21 & 97.5\% {\scriptsize(833)} & 91.8\% {\scriptsize(784)} \\
& D2 & 888 & -- & 14 & 98.4\% {\scriptsize(874)} & 93.4\% {\scriptsize(829)} \\
\cmidrule(lr){2-7}
& \textbf{All} & \textbf{1,742} & -- & \textbf{35} & \textbf{98\%} {\scriptsize(\textbf{1,707})} & \textbf{92.6\%} {\scriptsize(\textbf{1,613})} \\
\bottomrule
\end{tabular}
\end{wraptable} 

Conditioning on correct positive detection raises smell removal to 98.0\% and clean removal to 92.6\%. This refines the interpretation of RQ$_2$: many end-to-end failures reflect missed detections rather than unsuccessful patch generation. It also clarifies the role of task decomposition. D2 improves end-to-end results in part because it misses fewer positives than D1 (201 vs. 238), producing gains of 4.0 percentage points in SRR and 4.3 percentage points in CR. However, once the target smell is detected, patch quality is similar across configurations: D2 exceeds D1 by only 0.9 percentage points in conditional SRR and 1.6 percentage points in conditional CR. Thus, D2's refactoring benefit is primarily detection-driven, not evidence of systematically better patch generation.

\begin{finding}
End-to-end refactoring is detection-limited. Most Failed/No Patch outcomes are missed detections rather than patch-generation failures; when the target smell is correctly detected, clean removal rises from 74\% to 92.6\%.
\end{finding}

\textbf{Model-level outcomes show where D2's gains come from.}
Table~\ref{tab:rq2_clean_removal_by_model} shows that D2's aggregate refactoring gain is model-dependent. GPT-5.4-nano explains most of the reduction in detection misses, avoiding 33 of the 37 misses reduced by D2 (84 vs. 51); its end-to-end CR increases by 9.8 percentage points, while conditional CR remains similar (93.1\% vs. 92.0\%). Thus, for this model, decomposition mainly helps by detecting more positives. DeepSeek V4 Flash shows a different pattern: detection misses barely change (48 vs. 47), but conditional CR increases from 88.9\% to 94.2\%, indicating a gain after detection. Gemini 3.5 Flash and Claude Haiku 4.5 show smaller changes. Overall, D2's refactoring benefit is model-dependent: it is mainly detection-driven for GPT-5.4-nano, while the largest post-detection gain occurs for DeepSeek V4 Flash. Section~\ref{sec:rq3} tests whether these differences are statistically significant.

\begin{table}[ht!]
%\begin{wraptable}{r}{8cm}
\centering
\caption{Detection misses and Clean Removal (CR) under D1 and D2, by model. $\Delta$ = D2~$-$~D1 in percentage points, computed from unrounded values.}
\label{tab:rq2_clean_removal_by_model}
\scriptsize
\setlength{\tabcolsep}{3pt}
\renewcommand{\arraystretch}{1.08}
\begin{adjustbox}{max width=\linewidth}
\begin{tabular}{lrrrrrrrr}
\toprule
 & \multicolumn{2}{c}{\textbf{Detection misses}} & \multicolumn{3}{c}{\textbf{End-to-end CR}} & \multicolumn{3}{c}{\textbf{Conditional CR}} \\
\cmidrule(lr){2-3}\cmidrule(lr){4-6}\cmidrule(lr){7-9}
\textbf{Model} & D1 & D2 & D1 & D2 & $\Delta$ & D1 & D2 & $\Delta$ \\
\midrule
GPT-5.4-nano & 84 & 51 & 64.3\% & 74.1\% & +9.8 & 93.1\% & 92.0\% & -1.1 \\
Gemini 3.5 Flash & 63 & 57 & 70.1\% & 71.8\% & +1.7 & 91.3\% & 90.1\% & -1.2 \\
Claude Haiku 4.5 & 43 & 46 & 79.3\% & 80.7\% & +1.3 & 94.0\% & 96.9\% & +2.9 \\
DeepSeek V4 Flash & 48 & 47 & 73.3\% & 77.9\% & +4.7 & 88.9\% & 94.2\% & +5.3 \\
\midrule
\textbf{All models} & \textbf{238} & \textbf{201} & \textbf{71.8\%} & \textbf{76.1\%} & \textbf{+4.3} & \textbf{91.8\%} & \textbf{93.4\%} & \textbf{+1.6} \\
\bottomrule
\end{tabular}
\end{adjustbox}
%\end{wraptable} 
\end{table}

\textbf{Types of unnecessary edits.} For negative pairs, abstention is expected; any workflow modification is therefore unnecessary. Following Step~6 of Section~\ref{sec:dataCollection}, we classify modified negatives by their most behavior-relevant edit type. Since no modified negative output was classified as cosmetic-only, Table~\ref{tab:rq2_negative_edit_types} focuses on localized unnecessary edits and behavior-relevant edits. This distinction helps developers prioritize mitigation: improve abstention for plausible but unnecessary fixes, and add stronger safeguards for edits that may affect workflow execution.

\begin{table}[ht!]
\centering
\scriptsize
\caption{Types of unnecessary edits in modified ground-truth-negative pairs by design and model.}
\label{tab:rq2_negative_edit_types}
\setlength{\tabcolsep}{3.2pt}
\renewcommand{\arraystretch}{1.06}
\begin{adjustbox}{max width=\columnwidth}
\begin{tabular}{llrrr}
\toprule
\textbf{Edit type} & \textbf{Model}
& \textbf{D1}
& \textbf{D2}
& \textbf{Overall} \\
\midrule
\multirow{4}{*}{Localized but unnecessary edits}
& GPT-5.4-nano
& 57.14\%\,{\scriptsize(4/7)}
& 40.00\%\,{\scriptsize(6/15)}
& 45.45\%\,{\scriptsize(10/22)} \\
& Gemini 3.5 Flash
& 0.00\%\,{\scriptsize(0/3)}
& 12.50\%\,{\scriptsize(1/8)}
& 9.09\%\,{\scriptsize(1/11)} \\
& Claude Haiku 4.5
& 50.00\%\,{\scriptsize(7/14)}
& 60.00\%\,{\scriptsize(9/15)}
& 55.17\%\,{\scriptsize(16/29)} \\
& DeepSeek V4 Flash
& 50.00\%\,{\scriptsize(7/14)}
& 47.06\%\,{\scriptsize(8/17)}
& 48.39\%\,{\scriptsize(15/31)} \\
\midrule
\multirow{4}{*}{Behavior-relevant edits}
& GPT-5.4-nano
& 42.86\%\,{\scriptsize(3/7)}
& 60.00\%\,{\scriptsize(9/15)}
& 54.55\%\,{\scriptsize(12/22)} \\
& Gemini 3.5 Flash
& 100.00\%\,{\scriptsize(3/3)}
& 87.50\%\,{\scriptsize(7/8)}
& 90.91\%\,{\scriptsize(10/11)} \\
& Claude Haiku 4.5
& 50.00\%\,{\scriptsize(7/14)}
& 40.00\%\,{\scriptsize(6/15)}
& 44.83\%\,{\scriptsize(13/29)} \\
& DeepSeek V4 Flash
& 50.00\%\,{\scriptsize(7/14)}
& 52.94\%\,{\scriptsize(9/17)}
& 51.61\%\,{\scriptsize(16/31)} \\
\midrule
\textbf{Unnecessary edits} & \textbf{All models}
& \textbf{38} & \textbf{55} & \textbf{93} \\
\bottomrule
\end{tabular}
\end{adjustbox}
\end{table}

Across D1 and D2, 54.84\% of modified negative outputs are behavior-relevant (51/93), while 45.16\% are localized but unnecessary edits (42/93). This pattern is similar across configurations: behavior-relevant edits account for 52.63\% of modified negative outputs in D1 (20/38) and 56.36\% in D2 (31/55). Thus, when pipelines over-refactor, they often change executable or behavior-relevant workflow constructs rather than formatting, comments, or other cosmetic elements. Negative-pair outcomes also reveal smell-specific over-refactoring. As shown in Table~\ref{tab:rq2_refactoring_outcomes_by_smell}, unnecessary edits concentrate in \textit{MissingJobTimeout}, \textit{MissingConcurrencyGlobal}, and \textit{MissingConcurrencyBranch}, with UE rates of 44.44\%, 16.67\%, and 12.28\%, respectively. These cases, illustrated in Figure~\ref{fig:rq2_examples}(d), suggest that agents sometimes apply plausible fix templates even when the target smell is absent.

\begin{finding}
Negative pairs reveal model-specific abstention failures. All models produced unnecessary edits when the target smell was absent, although at different rates; overall, 54.84\% of these edits touched behavior-relevant workflow constructs.
\end{finding}

\subsection{Effectiveness--Overhead Trade-off of Task Decomposition (RQ$_3$)}
\label{sec:rq3}

To answer \textbf{RQ$_3$}, we compare D1 and D2 reusing RQ$_1$ detection outcomes and RQ$_2$ end-to-end clean removal, adding overhead measures: latency, token usage, and estimated API cost.

\textbf{Absolute overhead varies by model.} 
For each run, we aggregate provider-reported billable input and output tokens across all LLM calls and estimate API cost by multiplying them by the corresponding model-specific prices per one million tokens in Table~\ref{tab:evaluated_llms}. Token counts include reasoning or thinking tokens when reported as billable. Table~\ref{tab:rq3_absolute_overhead} reports median latency, token usage, and per-run estimated cost multiplied by 1,000 for readability. The paired tests report median within-pair D2--D1 differences, so the summaries may differ slightly.

\begin{table*}[ht!]
\centering
\scriptsize
\caption{Median operational overhead per \textit{(workflow, smell)} run by design and model. Cost reports median per-run estimated cost multiplied by 1,000. $\Delta$ = D2--D1.}
\label{tab:rq3_absolute_overhead}
\setlength{\tabcolsep}{3.2pt}
\renewcommand{\arraystretch}{1.05}
\begin{adjustbox}{max width=\columnwidth}
\begin{tabular}{l|rrr|rrr|rrr|rrr}
\toprule
\multirow{2}{*}{\textbf{Model}}
& \multicolumn{3}{c|}{\textbf{Latency (s)}}
& \multicolumn{3}{c|}{\textbf{Input tokens}}
& \multicolumn{3}{c|}{\textbf{Output tokens}}
& \multicolumn{3}{c}{\textbf{Median est. cost $\times$1K (USD)}} \\
\cmidrule(lr){2-4}
\cmidrule(lr){5-7}
\cmidrule(lr){8-10}
\cmidrule(lr){11-13}
& \textbf{D1} & \textbf{D2} & \textbf{$\Delta$}
& \textbf{D1} & \textbf{D2} & \textbf{$\Delta$}
& \textbf{D1} & \textbf{D2} & \textbf{$\Delta$}
& \textbf{D1} & \textbf{D2} & \textbf{$\Delta$} \\
\midrule
GPT-5.4-nano
& 14.5 & 11.1 & -3.4
& 3,888 & 2,655 & -1,233
& 2,210 & 1,546 & -664
& 3.63 & 2.47 & -1.16 \\

Gemini 3.5 Flash
& 17.9 & 9.7 & -8.2
& 12,743 & 2,941 & -9,802
& 2,660 & 1,186 & -1,474
& 42.04 & 14.71 & -27.33 \\

Claude Haiku 4.5
& 15.1 & 15.1 & +0.0
& 4,741 & 3,144 & -1,597
& 1,887 & 1,465 & -422
& 14.44 & 10.35 & -4.09 \\

DeepSeek V4 Flash
& 16.5 & 21.9 & +5.4
& 3,964 & 2,782 & -1,182
& 1,579 & 1,657 & +78
& 0.65 & 0.63 & -0.02 \\
\bottomrule
\end{tabular}
\end{adjustbox}
\end{table*}

We can observe in the table that D2 does not uniformly increase overhead. For GPT-5.4-nano and Gemini 3.5 Flash, D2 reduces median latency, token usage, and estimated cost. For Claude Haiku 4.5, D2 reduces token usage and estimated cost while keeping the same median latency. For DeepSeek V4 Flash, D2 reduces input tokens and keeps estimated cost nearly unchanged, but increases median latency and slightly increases output tokens. This pattern follows from D2's detection gate, which invokes refactoring only for predicted positives.

\textbf{Effectiveness--overhead trade-offs are model-dependent.}
Detection correctness, latency, and estimated cost use all 570 pairs; end-to-end clean removal uses only ground-truth-positive pairs whose D1 and D2 outputs are both assessable (255--268 pairs per model).
All effects are reported as D2--D1: for binary outcomes, $Eff=\mathrm{rate}_{D2}-\mathrm{rate}_{D1}$; for overhead outcomes, $Eff=\mathrm{median}_{i}(X_{i,D2}-X_{i,D1})$. Thus, positive binary effects favor D2, while negative overhead effects indicate that D2 reduces latency or estimated cost. Table~\ref{tab:rq3_paired_tests_across_configurations} reports the resulting comparisons.

\begin{table}[ht!]
\centering
\tiny
\caption{Paired D1--D2 tests for effectiveness and overhead. Binary effects are D2--D1 rate differences; overhead effects are median paired D2--D1 differences. $p$-values are Holm-corrected within each metric.}

\label{tab:rq3_paired_tests_across_configurations}
\begin{adjustbox}{max width=\columnwidth}
\begin{tabular}{ll|rrc|rrc|rrc|rrc}
\toprule
& & \multicolumn{3}{c|}{\textbf{Detection correctness}}
& \multicolumn{3}{c|}{\textbf{End-to-end clean removal}}
& \multicolumn{3}{c|}{\textbf{Latency}}
& \multicolumn{3}{c}{\textbf{Est. cost}} \\
\cmidrule(lr){3-5}
\cmidrule(lr){6-8}
\cmidrule(lr){9-11}
\cmidrule(lr){12-14}
\textbf{Comparison} & \textbf{Model}
& $p_{\mathrm{Holm}}$ & Eff. & Sig.
& $p_{\mathrm{Holm}}$ & Eff. & Sig.
& $p_{\mathrm{Holm}}$ & Eff. & Sig.
& $p_{\mathrm{Holm}}$ & Eff. & Sig. \\
\midrule
\multirow{4}{*}{D1 vs. D2}
& GPT-5.4-nano
& 0.015 & +0.04 & Yes
& 0.026 & +0.08 & Yes
& $<$0.001 & -3.5 & Yes
& $<$0.001 & -1.00 & Yes \\

& Gemini 3.5 Flash
& 1.000 & -0.01 & No
& 1.000 & +0.01 & No
& $<$0.001 & -3.5 & Yes
& $<$0.001 & -12.70 & Yes \\

& Claude Haiku 4.5
& 1.000 & 0.00 & No
& 1.000 & +0.00 & No
& $<$0.001 & -0.8 & Yes
& 0.933 & -3.58 & No \\

& DeepSeek V4 Flash
& 1.000 & 0.00 & No
& 0.242 & +0.05 & No
& $<$0.001 & +3.7 & Yes
& 0.933 & -0.04 & No \\
\bottomrule
\end{tabular}
\end{adjustbox}
\end{table}

Table~\ref{tab:rq3_paired_tests_across_configurations} shows that D2 significantly improves effectiveness only for GPT-5.4-nano, increasing detection correctness by 4 percentage points and end-to-end clean removal by 8 percentage points. However, the clean-removal gain is fragile: it is no longer significant when \textit{UnsafeDeployOnFork} is excluded (Section~\ref{sec:robustness_diagnostics}). The detection component of this gain is also smell-dependent: GPT-5.4-nano benefits from D2 mainly for \textit{SuboptimalCacheUsage}, \textit{MissingUploadCondition}, \textit{MissingForkGuardSchedule}, \textit{UnsafeArtifactUploadFork}, and \textit{SingleLangVersionCI}, whereas \textit{MissingJobTimeout} and concurrency-related smells show smaller gains or favor D1. For Gemini 3.5 Flash, Claude Haiku 4.5, and DeepSeek V4 Flash, D1 and D2 do not differ significantly in effectiveness. In contrast, overhead changes are more widespread, with latency differing significantly for all models. D2 is faster for GPT-5.4-nano, Gemini 3.5 Flash, and Claude Haiku 4.5, but slower for DeepSeek V4 Flash. The estimated cost decreases significantly only for GPT-5.4-nano and Gemini 3.5 Flash. Overall, task decomposition is an effectiveness--overhead trade-off rather than a default improvement.

\begin{finding}
Task decomposition is not uniformly better: D2 improves effectiveness significantly only for GPT-5.4-nano, while overhead changes vary by model.
\end{finding}

\subsection{Robustness and Diagnostic Analyses}
\label{sec:robustness_diagnostics}

Four diagnostic analyses were conducted to check if our conclusions depend on key design choices.

\textbf{\textit{GHA-based candidate retrieval.}}
The GHA analyzer should not be treated as a labeling oracle. Of the 1,416 GHA-flagged candidates inspected manually, annotators confirmed 902 and rejected 514; only $9/19$ smells reached $\geq 90\%$ confirmation, while $5/19$ were below $70\%$. GHA also missed positives: 366 of 1,310 GHA-unflagged candidates were manually labeled positive (27.9\%).
The lowest confirmation rate, 2.8\% for \textit{UnsafeDeployOnFork}, reflects a rule that flags any job without a repository guard, whether or not the job deploys. Some misses stem from narrower rules; for example, GHA does not flag branch references such as \texttt{@main} for \textit{UnpinnedActionVersion}. 
Thus, GHA is useful for reproducible retrieval and stratified sampling, but labels and reported effectiveness must rely on the manual ground truth. Moreover, the final natural ground truth keeps only candidates whose manual label agrees with the sampling stratum;
GHA matches every natural ground-truth label by construction, so we use it neither as a detection baseline for RQ$_1$ nor, since it does not generate patches, as a refactoring baseline for RQ$_2$.

\textbf{\textit{Balanced evaluation.}} Precision in our balanced 570-pair evaluation should not be interpreted as field precision under natural smell prevalence.
We therefore recomputed precision using candidate-pool prevalence estimated from the GHA-guided sampling pool and manual confirmation/miss rates among the 14,385 workflows eligible for Step~3.
Across the 19 smells, the estimated positive-pair prevalence was 29.92\% on average, ranging from 1.97\% to 93.06\%. Keeping predictions fixed, we recomputed precision as $\pi R / (\pi R + (1-\pi)\mathit{FPR})$, where $\pi$ is the estimated prevalence, $R$ is recall, and $\mathit{FPR}$ is the false-positive rate.
Averaged over the four models, prevalence-adjusted precision decreases from 95.38\% to 89.92\% for D1 and from 93.44\% to 85.89\% for D2, while recall remains 79.12\% and 82.37\%, respectively.
The main interpretation of RQ$_1$ remains:
D1 is more precise, while D2 recovers more smells. However, it nearly removes D2's aggregate F1 advantage, reducing the D2--D1 gap from 1.21 to 0.14 percentage points because D2's additional false positives in the balanced set weigh more heavily under lower positive-pair prevalence. This diagnostic reframes the precision under a less balanced prevalence assumption; it does not estimate the field precision because the candidate pool was GHA-guided and ranked rather than randomly sampled.

\textbf{\textit{Semi-synthetic \textit{UnsafeDeployOnFork}.}}
Because natural positives for \textit{UnsafeDeployOnFork} were scarce, we generated 49 semi-synthetic candidates from real workflows labeled negative for this smell; 45 were accepted as positives after manual validation. As a sanity check, GHA flagged 48/49 mutated workflows, including all accepted positives,
while every base workflow was GHA-unflagged by selection. However, GHA also flagged 3/4 rejected candidates, reinforcing that final labels must come from manual validation.
We repeated the main analyses after excluding this smell, i.e., its 30 pairs in the 570-pair set: 15 positives (10 semi-synthetic and 5 natural) and 15 natural negatives.
Reusing the same outputs and metadata, the exclusion changes detection F1 by $\leq 0.33$ percentage points and Clean Removal by $\leq 0.73$ percentage points per model and design.
We do not report changes in Unnecessary Edit, latency, or estimated cost because these outcomes do not depend on the semi-synthetic positive labels. 
The exclusion changes one RQ$_3$ significance conclusion: GPT-5.4-nano's end-to-end clean-removal gain remains similar in size (8.2 vs. 7.5 percentage points), but is no longer significant ($p_{\mathrm{Holm}}$: 0.026 $\rightarrow$ 0.070), because three of its 38 pairs with clean removal only under D2 are \textit{UnsafeDeployOnFork} pairs. Detection correctness, latency, and estimated-cost conclusions do not change. Thus, the main findings are not driven by this semi-synthetic smell, but the significance of GPT-5.4-nano's clean-removal gain rests on a narrow margin.

\textbf{Workflow-level feasibility diagnostic.}
Because \textsc{GHA-SmellFix} is target-conditioned, we ran a small diagnostic on eight real workflows to assess workflow-level smell-set recovery through exhaustive target-conditioned screening. Two annotators labeled all 19 smells per workflow, agreeing on 138/152 pairs ($\kappa = 0.759$); the workflows contain 40 present pairs across 16 smells. Using Claude Haiku 4.5, we ran D1 and D2 once per target smell, yielding 304 runs, and aggregated predicted-present smells into workflow-level sets. Both designs recovered all 40 annotated pairs (recall 100\%) and produced nine predicted-present pairs not annotated as present (precision 81.63\%); exact workflow-level matches occurred for 3/8 workflows under D1 and 2/8 under D2. We also inspected patches independently per target smell, without merging them into coordinated workflow-level refactorings. Both designs generated patches for all 40 present pairs, with 32/40 D1 and 33/40 D2 patches being valid smell-specific refactorings or requiring only external validation. On absent pairs, each design generated nine unnecessary patches, all affecting behavior-relevant constructs. Overall, this diagnostic supports the feasibility of workflow-level smell-set recovery and smell-specific patch generation, while reinforcing the need for reliable abstention and construct-aware validation.

\section{Study Implications}
\label{sec:implications}

\textbf{For researchers: evaluate CI/CD agents as maintenance systems, not only smell detectors.}
Existing rule-based work on GitHub Actions quality and security problems has mostly focused on detecting known configuration patterns~\cite{benedetti2022automatic,khatami2024catching,freitas2024gash}. Because we build on a developer-informed smell catalog, our results shift the focus from smell relevance to whether LLM-based pipelines can detect, refactor, and abstain safely when targeting these smells. 
A patch may remove the target smell while still introducing unexpected edits, potential behavior-impacting changes, or unnecessary edits, so CI/CD agents should be evaluated beyond detection and smell-removal rates.

\textbf{For tool builders: place LLM-generated workflow patches behind validation gates.} \textsc{GHA-SmellFix} can draft useful candidate patches, especially when targeting explicit constructs, but should not be treated as an autonomous refactoring tool. GitHub Actions workflows depend on behavior-relevant constructs such as permissions, triggers, runners, secrets, artifacts, deployments, reusable workflows, and repository-specific policies~\cite{github2026workflow,koishybayev2022characterizing}; small YAML edits can have execution-level consequences. Tool builders should combine LLM-generated patches with YAML and GitHub Actions structural checks, policy checks, construct-aware diff analysis, abstention checks, and repository-level validation for external facts like action SHAs or dependency versions. A practical deployment path is a hybrid static-analysis--agentic-LLM pipeline where rule-based scanners, like the GHA analyzer, provide first-level screening, while agentic LLM components generate candidate patches passing through construct-aware validation gates. Orchestration should be model- and smell-aware, since decomposition changed effectiveness and overhead in model-dependent ways.

\textbf{For developers and maintainers: treat generated patches as review candidates, not ready-to-merge fixes.} The studied designs produce useful candidate patches, but human review remains necessary. Maintainers should inspect whether a patch changes only expected workflow constructs, preserves intended triggers, permissions, runners, dependencies, and reusable-workflow calls, and uses contextually valid values. In practice, LLM agents are better positioned as review-support components for CI/CD configuration maintenance than as fully automatic refactoring systems.

\section{Threats to Validity}
\label{sec:threats}

\textbf{Construct validity.}
We derived our smell definitions from an existing GitHub Actions smell catalog~\cite{khatami2024catching} grounded in workflow evolution evidence and examined through a contribution study with maintainers. We operationalized the selected smells with explicit detection conditions, exclusion criteria, uncertainty rules, and counting units, and cross-checked them against official GitHub Actions~\cite{github2026workflow}. Still, some smells may depend on unobservable repository intent from workflow files, especially concerning fork guards, deployment behavior, concurrency semantics, reusable workflows, or workflow intent. Our refactoring assessment is construct-aware rather than runtime-semantic: using automatic checks, diffs, and manual inspection, we evaluate smell removal, YAML parsability, GitHub Actions structural validity, expected construct scope, and evident behavior-relevant changes. We do not execute workflows or verify runtime-dependent facts like commit hashes, package versions, credentials, external services, branch rules, or repository-specific policies. Thus, our results characterize structurally plausible candidate patches, not production-ready repairs.

\textbf{Internal validity.}
Our ground truth was manually annotated through independent inspection, consensus discussion, and agreement analysis with Cohen's Kappa~\cite{mchugh2012interrater,landis1977measurement}. Some annotators are co-authors, which may introduce confirmation bias. We mitigated this risk through explicit guidelines, pairwise annotation, consensus, fixed ground-truth labels before any agent run, and blinded inspection of refactoring outputs with respect to model and design identities. In addition, none of the annotators implemented the agentic pipelines or executed the experiments.

We used \textsc{GHA} analyzer~\cite{khatami2024catching} only for stratified sampling; final labels came from manual annotation. This reduces dependence on the initial heuristic tool, but may underrepresent positives missed by \textsc{GHA} analyzer or workflows with multiple co-occurring smells. The balanced ground truth supports paired comparisons between smells, models, and designs, but it does not estimate the prevalence of smells in the real-world. Thus, precision, recall, abstention, and unnecessary-edit rates should be interpreted under the sampled distribution, not as prevalence-adjusted field estimates.

For \textit{UnsafeDeployOnFork}, natural positives were scarce, so we used LLM-assisted semi-synthetic positives derived from real workflows by mutating only the target-smell construct. These cases were manually validated, and no synthetic negative pair was generated. Since semi-synthetic cases may differ from natural ones, we avoid conclusions that rely on this smell alone. Section~\ref{sec:robustness_diagnostics} shows that excluding this smell changes detection F1 and clean removal by less than one percentage point, but makes GPT-5.4-nano's clean-removal gain no longer statistically significant.

\textbf{Conclusion validity.}
We mitigate threats by using paired comparisons over the same \textit{(workflow, smell)} inputs, McNemar's exact test for binary outcomes~\cite{mcnemar1947note}, Wilcoxon signed-rank tests for paired overhead measures~\cite{wilcoxon1945individual}, and Holm correction for multiple comparisons~\cite{holm1979simple}. Still, the balanced shared set has limited per-smell sample size, and Gemini 3.5 Flash and DeepSeek V4 Flash were executed once, which may limit estimate precision. LLM executions may also vary due to provider-side nondeterminism, infrastructure changes, or model updates; we repeated GPT-5.4-nano and Claude Haiku 4.5 three times and limit stability conclusions to those models. Non-significant results should not be interpreted as evidence of equivalence~\cite{altman1995absence}. We report effect directions, significance results, generated outputs, and execution metadata in the replication package.

\textbf{External validity.}
We focus on GitHub Actions workflows sampled from a workflow-history dataset~\cite{cardoen2024dataset} and 19 configuration smells~\cite{khatami2024catching}.
The findings may not generalize to other CI/CD platforms, smell catalogs, proprietary workflows, larger workflows, repositories with many co-occurring smells, or teams with different CI/CD conventions. 
We evaluate target-conditioned detection and refactoring over explicit \textit{(workflow, smell)} pairs, not open-ended discovery, target-free refactoring, or coordinated multi-smell repair. Although Section~\ref{sec:robustness_diagnostics} reports a small workflow-level diagnostic, our main evaluation remains target-conditioned; larger-scale workflow-level screening and coordinated multi-smell refactoring require broader annotations and strategies for resolving interacting patches.
Although \textsc{GHA-SmellFix} is conceptually extensible to other CI/CD platforms, its smell definitions, prompts, construct mappings, and validation rules would need to be adapted to each platform's configuration language and execution semantics.

\section{Conclusion and Future Work}
\label{sec:conclusionAndFutureWork}

This paper presented \textsc{GHA-SmellFix}, a target-conditioned agentic pipeline for detecting and refactoring GitHub Actions configuration smells. We evaluated a joint detect-and-refactor design and a decomposed detect-then-refactor design with four LLMs on 570 shared \textit{(workflow, smell)} pairs sampled from a 1,710-pair manually curated ground truth covering 19 smells. Our results show that task decomposition helps selectively, but is not a default improvement. D2 modestly improves aggregate detection, increasing average F1 from 86.34\% to 87.55\% and recall from 79.12\% to 82.37\%, with the strongest gains for models that under-detect in D1. However, once the target smell is correctly detected, conditional patch quality is similar across D1 and D2. Thus, D2's main benefit is detection-driven rather than evidence of systematically better patch generation.

Refactoring results are promising but require safeguards. Across assessable positive outputs, 78.3\% remove the target smell and 74\% are clean removals; given correct detection, smell removal increases to 98\% and clean removal to 92.6\%. At the same time, smell removal alone is insufficient to characterize patch quality: some patches introduce unexpected edits or potential behavior-impacting changes, and negative pairs reveal 93 unnecessary edits, 54.84\% of which affect behavior-relevant workflow constructs. Overall, \textsc{GHA-SmellFix} supports review-oriented CI/CD maintenance, but requires validation and human review. Future work should incorporate richer repository context, external validators, sandboxed CI execution, broader CI/CD platforms and smell catalogs, and independent evaluation of generated patches with maintainers or external CI/CD experts.

\section*{Data Availability}
All datasets, prompts, outputs, and analysis scripts is available at: \hl{\url{https://github.com/reset-ufc/GHA-SmellFix-preprint}}.
\bibliographystyle{ACM-Reference-Format}
%\bibliography{acmart}
\bibliography{references}

@article{yang2024swe,
  title={Swe-agent: Agent-computer interfaces enable automated software engineering},
  author={Yang, John and Jimenez, Carlos E and Wettig, Alexander and Lieret, Kilian and Yao, Shunyu and Narasimhan, Karthik and Press, Ofir},
  journal={Advances in Neural Information Processing Systems},
  volume={37},
  pages={50528--50652},
  year={2024}
}

@inproceedings{hilton2017trade,
  title={Trade-offs in continuous integration: assurance, security, and flexibility},
  author={Hilton, Michael and Nelson, Nicholas and Tunnell, Timothy and Marinov, Darko and Dig, Danny},
  booktitle={Proceedings of the 2017 11th Joint Meeting on Foundations of Software Engineering},
  pages={197--207},
  year={2017}
}

@inproceedings{vassallo2020configuration,
  title={Configuration smells in continuous delivery pipelines: a linter and a six-month study on GitLab},
  author={Vassallo, Carmine and Proksch, Sebastian and Jancso, Anna and Gall, Harald C and Di Penta, Massimiliano},
  booktitle={Proceedings of the 28th ACM joint meeting on european software engineering conference and symposium on the foundations of software engineering},
  pages={327--337},
  year={2020}
}

@article{zampetti2020empirical,
  title={An empirical characterization of bad practices in continuous integration},
  author={Zampetti, Fiorella and Vassallo, Carmine and Panichella, Sebastiano and Canfora, Gerardo and Gall, Harald and Di Penta, Massimiliano},
  journal={Empirical Software Engineering},
  volume={25},
  number={2},
  pages={1095--1135},
  year={2020},
  publisher={Springer}
}

@inproceedings{vassallo2019automated,
  title={Automated reporting of anti-patterns and decay in continuous integration},
  author={Vassallo, Carmine and Proksch, Sebastian and Gall, Harald C and Di Penta, Massimiliano},
  booktitle={2019 IEEE/ACM 41st International Conference on Software Engineering (ICSE)},
  pages={105--115},
  year={2019},
  organization={IEEE}
}

@article{gallaba2018use,
  title={Use and misuse of continuous integration features: An empirical study of projects that (mis) use Travis CI},
  author={Gallaba, Keheliya and McIntosh, Shane},
  journal={IEEE Transactions on Software Engineering},
  volume={46},
  number={1},
  pages={33--50},
  year={2018},
  publisher={IEEE}
}

@inproceedings{bouzenia2025repairagent,
  title={Repairagent: An autonomous, llm-based agent for program repair},
  author={Bouzenia, Islem and Devanbu, Premkumar and Pradel, Michael},
  booktitle={2025 IEEE/ACM 47th International Conference on Software Engineering (ICSE)},
  pages={2188--2200},
  year={2025},
  organization={IEEE}
}

@inproceedings{wang2025openhands,
  title={Openhands: An open platform for ai software developers as generalist agents},
  author={Wang, Xingyao and Li, Boxuan and Song, Yufan and Xu, Frank F and Tang, Xiangru and Zhuge, Mingchen and Pan, Jiayi and Song, Yueqi and Li, Bowen and Singh, Jaskirat and others},
  booktitle={International Conference on Learning Representations},
  volume={2025},
  pages={65882--65919},
  year={2025}
}

@inproceedings{valenzuela2022evolution,
  title={Evolution of github action workflows},
  author={Valenzuela-Toledo, Pablo and Bergel, Alexandre},
  booktitle={2022 IEEE International Conference on Software Analysis, Evolution and Reengineering (SANER)},
  pages={123--127},
  year={2022},
  organization={IEEE}
}

@article{altman1995absence,
  title={Absence of evidence is not evidence of absence.},
  author={Altman, Douglas G and Bland, J Martin},
  journal={British Medical Journal},
  volume={311},
  number={7003},
  pages={485--486},
  year={1995},
  publisher={BMJ Publishing Group Ltd.}
}

@article{ye2021automated,
  title={Automated patch assessment for program repair at scale},
  author={Ye, He and Martinez, Matias and Monperrus, Martin},
  journal={Empirical Software Engineering},
  volume={26},
  number={2},
  pages={20},
  year={2021},
  publisher={Springer}
}

@inproceedings{beller2016analyzing,
  title={Analyzing the state of static analysis: A large-scale evaluation in open source software},
  author={Beller, Moritz and Bholanath, Radjino and McIntosh, Shane and Zaidman, Andy},
  booktitle={2016 IEEE 23rd International Conference on Software Analysis, Evolution, and Reengineering (SANER)},
  volume={1},
  pages={470--481},
  year={2016},
  organization={IEEE}
}

@inproceedings{johnson2013don,
  title={Why don't software developers use static analysis tools to find bugs?},
  author={Johnson, Brittany and Song, Yoonki and Murphy-Hill, Emerson and Bowdidge, Robert},
  booktitle={2013 35th International Conference on Software Engineering (ICSE)},
  pages={672--681},
  year={2013},
  organization={IEEE}
}

@misc{github2026workflow,
  author       = {{GitHub}},
  title        = {Workflow syntax for GitHub Actions},
  year         = {2026},
  howpublished = {\url{https://docs.github.com/actions/using-workflows/workflow-syntax-for-github-actions}},
  note         = {Accessed: 2026-04-17}
}

@article{fawcett2006introduction,
  title={An introduction to ROC analysis},
  author={Fawcett, Tom},
  journal={Pattern Recogn. Lett.},
  volume={27},
  number={8},
  pages={861--874},
  year={2006}
}

@inproceedings{freitas2024gash,
  title={GASH--The GitHub Actions Smell Hunter},
  author={Freitas, Matheus B and Rocha, Lincoln S},
  booktitle={Workshop de Visualiza{\c{c}}{\~a}o, Evolu{\c{c}}{\~a}o e Manuten{\c{c}}{\~a}o de Software (VEM)},
  pages={103--112},
  year={2024},
  organization={SBC}
}

@inproceedings{cardoen2024dataset,
author = {Cardoen, Guillaume and Mens, Tom and Decan, Alexandre},
title = {A dataset of GitHub Actions workflow histories},
year = {2024},
isbn = {9798400705878},
publisher = {Association for Computing Machinery},
address = {New York, NY, USA},
url = {https://doi.org/10.1145/3643991.3644867},
doi = {10.1145/3643991.3644867},
booktitle = {Proceedings of the 21st International Conference on Mining Software Repositories},
pages = {677–681},
numpages = {5},
location = {Lisbon, Portugal},
series = {MSR '24}
}

@inproceedings{khatami2024catching,
  title={Catching smells in the act: A github actions workflow investigation},
  author={Khatami, Ali and Willekens, C{\'c}dric and Zaidman, Andy},
  booktitle={2024 IEEE International Conference on Source Code Analysis and Manipulation (SCAM)},
  pages={47--58},
  year={2024},
  organization={IEEE}
}

@article{derouiche2025agentic,
  title={Agentic ai frameworks: Architectures, protocols, and design challenges},
  author={Derouiche, Hana and Brahmi, Zaki and Mazeni, Haithem},
  journal={arXiv preprint arXiv:2508.10146},
  year={2025}
}

@article{jiang2026survey,
  title={A survey on large language models for code generation},
  author={Jiang, Juyong and Wang, Fan and Shen, Jiasi and Kim, Sungju and Kim, Sunghun},
  journal={ACM Transactions on Software Engineering and Methodology},
  volume={35},
  number={2},
  pages={1--72},
  year={2026},
  publisher={ACM New York, NY}
}

@inproceedings{liu2024agentbench,
  title={Agentbench: Evaluating llms as agents},
  author={Liu, Xiao and Yu, Hao and Zhang, Hanchen and Xu, Yifan and Lei, Xuanyu and Lai, Hanyu and Gu, Yu and Ding, Hangliang and Men, Kaiwen and Yang, Kejuan and others},
  booktitle={International Conference on Learning Representations},
  volume={2024},
  pages={52989--53046},
  year={2024}
}

@article{gheyi2026foundation,
  title={Foundation models as oracles for refactoring correctness detection},
  author={Gheyi, Rohit and Melo, Rian and Oliveira, Jonhnanthan and Ribeiro, M{\'a}rcio and Fonseca, Baldoino},
  journal={Empirical Software Engineering},
  volume={32},
  number={1},
  pages={32},
  year={2027},
  publisher={Springer}
}

@article{wilcoxon1945individual,
  title={Individual comparisons by ranking methods},
  author={Wilcoxon, Frank},
  journal={Biometrics bulletin},
  volume={1},
  number={6},
  pages={80--83},
  year={1945},
  publisher={JSTOR}
}

@article{holm1979simple,
  title={A simple sequentially rejective multiple test procedure},
  author={Holm, Sture},
  journal={Scandinavian journal of statistics},
  pages={65--70},
  year={1979},
  publisher={JSTOR}
}

@inproceedings{benedetti2022automatic,
  title={Automatic security assessment of github actions workflows},
  author={Benedetti, Giacomo and Verderame, Luca and Merlo, Alessio},
  booktitle={Proceedings of the 2022 ACM Workshop on Software Supply Chain Offensive Research and Ecosystem Defenses},
  pages={37--45},
  year={2022}
}

@inproceedings{bardsiri2026github,
  title={On the GitHub Actions Language: Usage, Evolution, and Workflow Reliability},
  author={Bardsiri, Aref Talebzadeh and Decan, Alexandre and Mens, Tom},
  booktitle={42nd 2026 IEEE International Conference on Software Maintenance and Evolution (ICSME)},
  year={2026},
  url={https://arxiv.org/abs/2605.26825}, 
}

@article{patcas2026evaluation,
  title={An evaluation study of large language models for addressing code quality issues},
  author={Patcas, Rares and Motogna, Simona},
  journal={Empirical Software Engineering},
  volume={31},
  number={5},
  pages={118},
  year={2026},
  publisher={Springer}
}

@article{hou2024large,
  title={Large language models for software engineering: A systematic literature review},
  author={Hou, Xinyi and Zhao, Yanjie and Liu, Yue and Yang, Zhou and Wang, Kailong and Li, Li and Luo, Xiapu and Lo, David and Grundy, John and Wang, Haoyu},
  journal={ACM Transactions on Software Engineering and Methodology},
  volume={33},
  number={8},
  pages={1--79},
  year={2024},
  publisher={ACM New York, NY}
}

@article{wang2024survey,
  title={A survey on large language model based autonomous agents},
  author={Wang, Lei and Ma, Chen and Feng, Xueyang and Zhang, Zeyu and Yang, Hao and Zhang, Jingsen and Chen, Zhiyuan and Tang, Jiakai and Chen, Xu and Lin, Yankai and others},
  journal={Frontiers of Computer Science},
  volume={18},
  number={6},
  pages={186345},
  year={2024},
  publisher={Springer}
}

@inproceedings{decan2022use,
  title={On the use of github actions in software development repositories},
  author={Decan, Alexandre and Mens, Tom and Mazrae, Pooya Rostami and Golzadeh, Mehdi},
  booktitle={2022 IEEE International Conference on Software Maintenance and Evolution (ICSME)},
  pages={235--245},
  year={2022},
  organization={IEEE}
}

@inproceedings{kalliamvakou2014promises,
  title={The promises and perils of mining github},
  author={Kalliamvakou, Eirini and Gousios, Georgios and Blincoe, Kelly and Singer, Leif and German, Daniel M and Damian, Daniela},
  booktitle={Proceedings of the 11th working conference on mining software repositories},
  pages={92--101},
  year={2014}
}

@article{pickerill2020phantom,
  title={PHANTOM: Curating GitHub for engineered software projects using time-series clustering},
  author={Pickerill, Peter and Jungen, Heiko Joshua and Ochodek, Miros{\l}aw and Ma{\'c}kowiak, Micha{\l} and Staron, Miroslaw},
  journal={Empirical Software Engineering},
  volume={25},
  number={4},
  pages={2897--2929},
  year={2020},
  publisher={Springer}
}

@article{munaiah2017curating,
  title={Curating github for engineered software projects},
  author={Munaiah, Nuthan and Kroh, Steven and Cabrey, Craig and Nagappan, Meiyappan},
  journal={Empirical Software Engineering},
  volume={22},
  number={6},
  pages={3219--3253},
  year={2017},
  publisher={Springer}
}

@inproceedings{koishybayev2022characterizing,
  title={Characterizing the security of github $\{$CI$\}$ workflows},
  author={Koishybayev, Igibek and Nahapetyan, Aleksandr and Zachariah, Raima and Muralee, Siddharth and Reaves, Bradley and Kapravelos, Alexandros and Machiry, Aravind},
  booktitle={31st USENIX Security Symposium (USENIX Security 22)},
  pages={2747--2763},
  year={2022}
}

@article{fares2026unpacking,
  title={Unpacking Security Scanners for GitHub Actions Workflows},
  author={Fares, Madjda and Gamage, Yogya and Baudry, Benoit},
  journal={arXiv preprint arXiv:2601.14455},
  year={2026}
}

@inproceedings{bouzenia2024resource,
  title={Resource usage and optimization opportunities in workflows of github actions},
  author={Bouzenia, Islem and Pradel, Michael},
  booktitle={Proceedings of the 46th IEEE/ACM International Conference on Software Engineering},
  pages={1--12},
  year={2024}
}

@article{mazrae2026empirical,
  title={An empirical study of the evolution of GitHub Actions workflows},
  author={Mazrae, Pooya Rostami and Decan, Alexandre and Mens, Tom and Wessel, Mairieli},
  journal={Journal of Systems and Software},
  pages={112824},
  year={2026},
  publisher={Elsevier}
}

@article{driscoll1996robustness,
  title={Robustness of the ANOVA and Tukey-Kramer statistical tests},
  author={Driscoll, Wade C},
  journal={Computers \& Industrial Engineering},
  volume={31},
  number={1-2},
  pages={265--268},
  year={1996},
  publisher={Elsevier}
}

@article{mcnemar1947note,
  title={Note on the sampling error of the difference between correlated proportions or percentages},
  author={McNemar, Quinn},
  journal={Psychometrika},
  volume={12},
  number={2},
  pages={153--157},
  year={1947},
  publisher={Springer}
}

@article{shinn2023reflexion,
  title={Reflexion: Language agents with verbal reinforcement learning},
  author={Shinn, Noah and Cassano, Federico and Gopinath, Ashwin and Narasimhan, Karthik and Yao, Shunyu},
  journal={Advances in neural information processing systems},
  volume={36},
  pages={8634--8652},
  year={2023}
}

@article{melo2025agentic,
  title={Agentic LMs: Hunting Down Test Smells},
  author={Melo, Rian and Sim{\~o}es, Pedro and Gheyi, Rohit and d’Amorim, Marcelo and Ribeiro, M{\'a}rcio and Soares, Gustavo and Almeida, Eduardo and Soares, Elvys},
  journal={IEEE Software},
  year={2026},
  publisher={IEEE},
  volume       = {43},
  number       = {1},
  pages        = {32--40},
}

@article{schulhoff2024prompt,
  title={The prompt report: A systematic survey of prompt engineering techniques},
  author={Schulhoff, Sander and Ilie, Michael and Balepur, Nishant and Kahadze, Konstantine and Liu, Amanda and Si, Chenglei and Li, Yinheng and Gupta, Aayush and Han, HyoJung and Schulhoff, Sevien and others},
  journal={arXiv preprint arXiv:2406.06608},
  year={2024}
}

@article{landis1977measurement,
  title={The measurement of observer agreement for categorical data},
  author={Landis, J Richard and Koch, Gary G},
  journal={biometrics},
  pages={159--174},
  year={1977},
  publisher={JSTOR}
}

@article{mchugh2012interrater,
  title={Interrater reliability: the kappa statistic},
  author={McHugh, Mary L},
  journal={Biochemia medica},
  volume={22},
  number={3},
  pages={276--282},
  year={2012},
  publisher={Hrvatsko dru{\v{s}}tvo za medicinsku biokemiju i laboratorijsku medicinu}
}

@article{caldiera1994goal,
  title={Goal question metric paradigm},
  author={Caldiera, Victor R Basili-Gianluigi and Rombach, H Dieter},
  journal={Encyclopedia of software engineering},
  volume={1},
  number={528-532},
  pages={6},
  year={1994}
}

@article{xia2025agentless,
  title={Demystifying llm-based software engineering agents},
  author={Xia, Chunqiu Steven and Deng, Yinlin and Dunn, Soren and Zhang, Lingming},
  journal={Proceedings of the ACM on Software Engineering},
  volume={2},
  number={FSE},
  pages={801--824},
  year={2025},
  publisher={ACM New York, NY, USA}
}

@article{sun2026test,
  title={Test Code Review in the Era of GitHub Actions: A Replication Study},
  author={Sun, Hui and Wu, Yinan and Assun{\c{c}}{\~a}o, Wesley KG and Stolee, Kathryn T},
  journal={arXiv preprint arXiv:2603.15935},
  year={2026}
}

\end{document}